\documentclass[english,aps,pre,preprint,nofootinbib]{revtex4-1}
\usepackage[T1]{fontenc}
\usepackage[latin9]{inputenc}
\usepackage{color}
\usepackage{babel}
\usepackage{array}
\usepackage{bm}
\usepackage{multirow}
\usepackage{amsmath}
\usepackage{amssymb}
\usepackage{graphicx}
\usepackage{esint}
\usepackage[unicode=true,pdfusetitle,
 bookmarks=true,bookmarksnumbered=false,bookmarksopen=false,
 breaklinks=false,pdfborder={0 0 1},backref=false,colorlinks=false]
 {hyperref}

\makeatletter

\providecommand{\tabularnewline}{\\}

\usepackage{bm}
\usepackage{color}

\newcommand{\CC}[1]{{}}
\newcommand{\bS}{\begin{subequations}}
\newcommand{\eS}{\end{subequations}}

\DeclareMathOperator{\Pf}{Pf}
\DeclareMathOperator{\Tr}{Tr}
\DeclareMathOperator{\diag}{\mathbf{diag}}
\DeclareMathOperator{\sign}{sign}
\DeclareMathOperator{\QP}{\Pi}

\DeclareMathOperator{\artanh}{artanh}
\DeclareMathOperator{\arcoth}{arcoth}

\newcommand{\cI}[1]{(I.#1)}

\makeatother

\begin{document}
\global\long\def\mat#1{\bm{\mathrm{#1}}}

\CC{\global\long\def\Pf{\mathrm{Pf}}
\global\long\def\paraNO{\parallel}
\global\long\def\Tr{\mathrm{Tr}}
\global\long\def\diag{\mat{\mathrm{diag}}}
\global\long\def\sign{\mathrm{sign}}
\global\long\def\QP{\Pi}

\global\long\def\artanh{\mathrm{artanh}}
\global\long\def\arcoth{\mathrm{arcoth}}
\global\long\def\arccot{\mathrm{arccot}}
 see preamble }

\global\long\def\Def{\mathcal{\equiv}}
\let\Def=\equiv

\global\long\def\T#1{\bar{#1}}
\global\long\def\Tc{T_{\mathrm{c}}}

\global\long\def\L{\leftrightarrow}
\global\long\def\M{\updownarrow}
\global\long\def\gm{\circ}

\global\long\def\LL{L_{\L}}
\global\long\def\LM{L_{\M}}
\global\long\def\xL{x_{\L}}
\global\long\def\xM{x_{\M}}

\global\long\def\zc{z_{\mathrm{c}}}
\global\long\def\ii{\mathrm{i}}
\global\long\def\ee{\mathrm{e}}
\global\long\def\dd{\mathrm{d}}

\global\long\def\TM#1{\mat{\mathcal{T}}_{\!\!#1}}

\global\long\def\S{\vphantom{\T{\mathbf{s}}}\mathbf{s}}
\global\long\def\Sc{\T{\mathbf{s}}}
\global\long\def\So{\mathbf{o}}
\global\long\def\Se{\mathbf{e}}

\global\long\def\strip{\mathrm{strip}}
\global\long\def\Zstrip{Z_{\strip}}
\global\long\def\Fstrip{F_{\strip}}

\global\long\def\Zinf{Z_{\infty}}
\global\long\def\Finf{F_{\infty}}

\global\long\def\Zres{Z_{\infty}^{\mathrm{res}}}
\global\long\def\Fres{F_{\infty}^{\mathrm{res}}}

\global\long\def\Zsres{Z_{\strip}^{\mathrm{res}}}
\global\long\def\Fsres{F_{\strip}^{\mathrm{res}}}

\global\long\def\fb{f_{\mathrm{b}}}
\global\long\def\fs{f_{\mathrm{s}}}
\global\long\def\fc{f_{\mathrm{c}}}
\global\long\def\FC{\mathcal{F}}

\global\long\def\fbs{f_{\mathrm{b,s}}}
\global\long\def\fbsres{f_{\mathrm{b,s}}^{\mathrm{res}}}
\global\long\def\Fsc{F_{\mathrm{s,c}}}
\global\long\def\Fscres{F_{\mathrm{s,c}}^{\mathrm{res}}}

\global\long\def\dom#1{\hat{#1}}
\global\long\def\Res#1#2{\mathrm{Res}_{#1}#2}

\global\long\def\scal#1{\tilde{#1}}
\global\long\def\scalZ{\Sigma}
\global\long\def\scalF{\Psi}
\global\long\def\Thetaoo{\Theta^{\mathrm{(oo)}}}

\global\long\def\mystack#1#2{{{\scriptstyle #1}\atop {\scriptstyle #2}}}

\textcolor{cyan}{}\global\long\def\weg#1{{\color{cyan}\cancel{#1}}}

\title{The square lattice Ising model on the rectangle\\
 II: Finite-size scaling limit}

\author{Alfred Hucht}

\affiliation{Faculty for Physics, University of Duisburg-Essen, 47058 Duisburg,
Germany}
\begin{abstract}
Based on the results published recently \cite{Hucht16a}, the universal
finite-size contributions to the free energy of the square lattice
Ising model on the $L\times M$ rectangle, with open boundary conditions
in both directions, are calculated exactly in the finite-size scaling
limit $L,M\to\infty$, $T\to\Tc$, with fixed temperature scaling
variable $x\propto(T/\Tc-1)M$ and fixed aspect ratio $\rho\propto L/M$.
We derive exponentially fast converging series for the related Casimir
potential and Casimir force scaling functions. At the critical point
$T=\Tc$ we confirm predictions from conformal field theory \cite{CardyPeschel88,KlebanVassileva91}.
The presence of corners and the related corner free energy has dramatic
impact on the Casimir scaling functions and leads to a logarithmic
divergence of the Casimir potential scaling function at criticality.
\end{abstract}
\maketitle
\tableofcontents{}

\vfill{}

\section{Introduction}

In the first part of this work \cite{Hucht16a}, denoted I in the
following, we computed the partition function $Z$ of the two-dimensional
Ising model on the $L\times M$ rectangle with open boundary conditions
in both directions and with anisotropic reduced couplings $K^{\L}$
and $K^{\M}$ in horizontal and vertical direction, in units of $k_{\mathrm{B}}T$
with Boltzmann constant $k_{\mathrm{B}}$, at arbitrary temperatures
below and above the critical point. This second part is devoted to
the finite-size scaling (FSS) behavior near criticality. 

\begin{figure}
\begin{centering}
\includegraphics[width=0.9\textwidth]{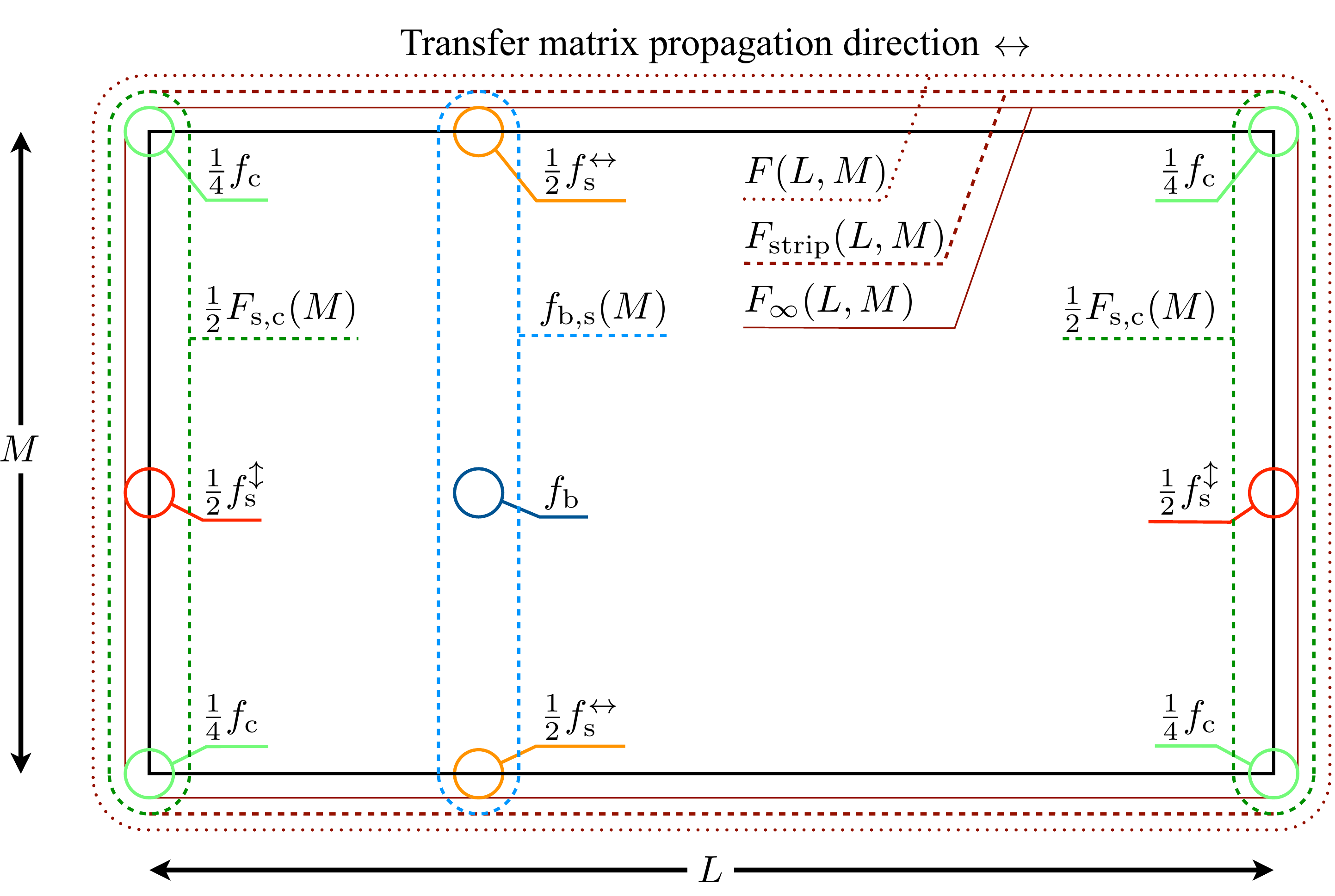}
\par\end{centering}
\caption{Sketch of the decomposition of the total free energy $F(L,M)$ (dotted
line) into the different constituents. The black rectangle is the
system, and the other solid lines denote infinite volume contributions
solely dependent on temperature, while the dashed lines symbolize
free energy parts containing residual finite-size contributions. \label{fig:System}}
\end{figure}

We first recall the main results of part I in terms of the size dependent
reduced free energy $F(L,M)=-\log Z$. From  \cI{87} we get the total
free energy of the considered model,
\begin{align}
F(L,M)={} & \underbrace{-\frac{L}{2}\left[M\log\left(\frac{2}{-z_{-}}\right)+\sum_{\mu=1}^{M}\dom{\gamma}_{\mu}\right]}_{L\fbs(M)}\nonumber \\
 & \underbrace{{}-\frac{1}{2}\log\Bigg[\left(\frac{2}{t_{-}z_{-}}\right)^{\frac{M^{2}}{2}}d_{\So,\Se}^{2}\,\prod_{\mu=1}^{M}\frac{(t_{+}z_{+}-\dom{\lambda}_{\mu,+})^{2}-t_{-}^{2}z_{-}^{2}}{M\dom{\lambda}_{\mu,-}^{2}+z_{+}\dom{\lambda}_{\mu,+}-t_{+}}\,\frac{z\dom{\lambda}_{\mu,-}}{v_{\mu}}\Bigg]}_{\Fsc(M)}\nonumber \\
 & \underbrace{{\vphantom{\Big[}}-\log\det(\mat 1+\mat Y)}_{\Fsres(L,M)},\label{eq:F}
\end{align}
where $\dom{\lambda}_{\mu}=\ee^{\dom{\gamma}_{\mu}}>1$ are the $M$
dominant eigenvalues of the $2M\times2M$ transfer matrix $\TM 2$
\cI{27}, given by the positive zeroes of the characteristic polynomial
$P_{M}(\varphi)$ \cI{45}. Alternatively, $\dom{\lambda}_{\mu,+}=\cosh\dom{\gamma}_{\mu}$
are the $M$ eigenvalues of $\TM +$ \cI{30a}. $z=\tanh K^{\L}$
and $t=\exp(-2K^{\M})$ parametrize the two couplings, and $a_{\pm}\Def{\textstyle \frac{1}{2}}(a\pm a^{-1})$
is a handy shortcut from \cI{20}. For a definition of the other quantities
$d_{\So,\Se}$ \cI{70b}, $v_{\mu}$ \cI{87a} and the $M/2\times M/2$
residual matrix $\mat Y$ \cI{87b} the reader is referred to part
I. In  (\ref{eq:F}) we decomposed the leading term $\Fstrip$ from
\cI{92} into two parts, 
\begin{equation}
\Fstrip(L,M)=L\,\fbs(M)+\Fsc(M),\label{eq:F_strip}
\end{equation}
one proportional to $L$ with contributions from the bulk and from
the two horizontal ($\L$) surfaces, and one with the remaining contributions
from the two vertical ($\M$) surfaces and from the four corners,
see figure \ref{fig:System}. These terms have been analyzed in great
detail by R.~J.~Baxter recently\footnote{The couplings $(z,t)$ are denoted $(t^{*},u^{*})$ in \cite{Baxter16},
where $z^{*}=\frac{1-z}{1+z}$ is the dual of $z$.} \cite{Baxter16}. 

For a detailed discussion of the different free energy contributions
we also recall the definition of the (total) residual free energy,
or Casimir potential, $\Fres$ \cI{89}, which is responsible for
nontrivial finite-size effects such as the critical Casimir effect
\cite{FisherdeGennes78}, \bS \label{eq:F-res_and_F_inf}
\begin{equation}
\Fres(L,M)\Def F(L,M)-\Finf(L,M),\label{eq:F-res}
\end{equation}
where the infinite volume contribution $\Finf$ can be decomposed
into bulk, surface and corner contributions according to \cI{90}
\begin{equation}
\Finf(L,M)\Def LM\fb+L\fs^{\L}+M\fs^{\M}+\fc\label{eq:F_inf}
\end{equation}
\eS{}for our rectangular geometry. The bulk free energy per spin
$\fb$, the surface free energies per surface spin pair $\fs^{\delta}$,
with direction $\delta\in\{\L,\M\}$, and the corner free energy $\fc$
are defined in the thermodynamic limit $L,M\to\infty$ and do not
depend on $L$ or $M$. While $\fb$ and $\fs^{\delta}$ are known
since a long time from the seminal works of Onsager \cite{Onsager44}
and McCoy \& Wu \cite{McCoyWu73}, the corner free energy $\fc$ below
$\Tc$ was derived only recently by Baxter\footnote{A constant term $-\log2$ attributed to $\fc^{<}$ in \cite{Baxter16,Hucht16a}
stems from the broken symmetry below $\Tc$ and has to be moved from
$\Finf$ to $\Fres$, see section~\ref{subsec:Potential}.} \cite{Baxter16}, confirming a conjectured product formula by Vernier
\& Jacobsen \cite{VernierJacobsen12}. The corresponding product formula
for temperatures above $\Tc$ was given in \cI{A7d}. The logarithmic
divergence $\fc\simeq\frac{1}{8}\log|1-T/\Tc|+\mathcal{O}(1)$ detailed
in appendix \ref{sec:q-expansion} will lead to a considerable complification
of the FSS analysis, as shown below.

Comparing (\ref{eq:F}) with (\ref{eq:F-res_and_F_inf}), we first
focus on the $L$ dependent terms. The free energy per row $\fbs$
is the dominant outcome of one application of the transfer matrix
$\TM 2$. It can be further decomposed, 
\begin{equation}
\fbs(M)=M\fb+\fs^{\L}+\fbsres(M),\label{eq:f_bs}
\end{equation}
and contains a residual contribution at finite $M$ that is equal
to the leading large-$L$ contribution to $\Fres$ from \cI{94},
\begin{equation}
\fbsres(M)\Def\lim_{L\to\infty}L^{-1}\Fres(L,M).\label{eq:f_bs-res}
\end{equation}
Inserting (\ref{eq:f_bs}) into (\ref{eq:F}), 
\begin{equation}
F(L,M)=L\big[M\fb+\fs^{\L}+\fbsres(M)\big]+\Fsc(M)+\Fsres(L,M),\label{eq:F-inserted}
\end{equation}
and matching with (\ref{eq:F-res_and_F_inf}), we can eliminate the
leading $L$-dependent terms and find three contributions to the total
residual free energy (\ref{eq:F-res}), 
\begin{equation}
\Fres(L,M)=L\fbsres(M)+\Fscres(M)+\Fsres(L,M),\label{eq:F-res-sum}
\end{equation}
where we defined the residual surface-corner contribution
\begin{equation}
\Fscres(M)\Def\Fsc(M)-\big[M\fs^{\M}+\fc\big],\label{eq:F_sc-res}
\end{equation}
which is independent of $L$. Equation (\ref{eq:F-res-sum}) shows
that the residual free energy can be decomposed very similar to (\ref{eq:F-res_and_F_inf}).

We now turn to the critical Casimir force per area $M$, that is defined
as $L$-derivative of the total residual free energy (\ref{eq:F-res}),
\begin{equation}
\FC(L,M)\Def-\frac{1}{M}\frac{\partial}{\partial L}\Fres(L,M),\label{eq:FC}
\end{equation}
and by (\ref{eq:F-res-sum}) decomposes into two contributions,
\begin{equation}
\FC(L,M)=\underbrace{{}-\frac{1}{M}f_{\mathrm{b,s}}^{\mathrm{res}}(M)}_{\FC_{\mathrm{b,s}}(M)}\underbrace{{}-\frac{1}{M}\frac{\partial}{\partial L}\Fsres(L,M)}_{\FC_{\strip}(L,M)},\label{eq:FC-sum}
\end{equation}
where the first one $\FC_{\mathrm{b,s}}$ is already known from the
stripe geometry $L/M\to\infty$. Note that the surface-corner contribution
$\Fscres$ drops out in the $L$-derivative as expected \cite{DieGruHaHuRuSch12,DieGruHaHuRuSch14a},
which renders the analysis of the Casimir force $\FC$ simpler than
the analysis of the Casimir potential $\Fres$.

With these definitions, we now summarize the FSS theory for the Casimir
potential and Casimir force, and introduce the corresponding universal
FSS functions.

\section{Finite-size scaling theory}

In this chapter we will formulate the finite-size scaling theory for
the residual free energy, or critical Casimir potential, as well as
for the critical Casimir force per area in the general case of a $d$-dimensional
weakly anisotropic system\footnote{For a discussion of weakly vs. strongly anisotropic critical behavior
see, e.\,g., \cite{Hucht02a}.} with size $V=\LL\LM^{d-1}$ and different couplings $K^{\L}$ and
$K^{\M}$. We use $\L$ for the direction parallel to the force and
$\M$ for all other directions, and we rewrite $\LL\Def L$ and $\LM\Def M$
in this and in the following chapter. When we later apply this theory
to our model, we will let the transfer matrix $\TM{}$ propagate parallel
to the Casimir force, as the calculation of the force requires the
derivative of the residual free energy with respect to $L$. 

Near criticality, the anisotropic couplings $K^{\L}$ and $K^{\M}$
lead to weakly anisotropic critical behavior, characterized by a weakly
anisotropic bulk correlation length\footnote{``$\simeq$'' denoted ``asymptotically equal''}
\begin{equation}
\xi_{\infty}^{\delta}(\tau)\stackrel{\tau>0}{\simeq}\xi_{+}^{\delta}\tau^{-\nu},\label{eq:xi_bulk}
\end{equation}
where $\tau=T/\Tc-1$ denotes the reduced temperature, and $\xi_{+}^{\delta}$
denotes the correlation length amplitude in direction $\delta\in\{\L,\M\}$
above criticality. The correlation length exponent is $\nu=1$ for
the $2d$ Ising model. Finite-size scaling theory predicts that near
the critical point $\tau\to0$ and for $L_{\delta}\to\infty$ with
fixed geometric aspect ratio $r=\LL/\LM$, the residual free energy
$\Fres$ (\ref{eq:F-res}) only depends on the two length ratios $\LL/\xi_{\infty}^{\L}(\tau)$
and $\LM/\xi_{\infty}^{\M}(\tau)$ \cite{LandauSwendsen84,Indekeu86,Hucht02a}.
These two ratios can be combined to a \emph{reduced aspect ratio}
\bS \label{eq:x,rho_0}
\begin{equation}
\rho\Def\frac{\LL\,\xi_{+}^{\M}}{\LM\,\xi_{+}^{\L}}\stackrel{\tau>0}{\simeq}\frac{\LL\,\xi_{\infty}^{\M}(\tau)}{\LM\,\xi_{\infty}^{\L}(\tau)},\label{eq:rho_0}
\end{equation}
which does not depend on temperature and encodes both the system shape
as well as the coupling anisotropy \cite{Hucht02a}. 

Approaching the critical point, the critical correlations are bounded
by the smallest length ratio in the system. For the given geometry
with arbitrary reduced aspect ratio $\rho$ this leads to three different
possible choices for the temperature scaling variable $x$ \cite{HuchtGruenebergSchmidt11,HobrechtHucht16a}:
If $0\leq\rho\lesssim1$ ($1\lesssim\rho\leq\infty$), the correlations
are limited by $\LL$ ($\LM$), while for arbitrary finite aspect
ratio $0<\rho<\infty$ the geometric mean $L_{\gm}\Def V^{1/d}$ can
be used as relevant length, avoiding the preference for one direction
$\delta$. In all three cases, the temperature scaling variable is
given by 
\begin{equation}
x_{\delta}\equiv\tau\left(\frac{L_{\delta}}{\xi_{+}^{\delta}}\right)^{\!\frac{1}{\nu}}\stackrel{\tau>0}{\simeq}\left(\frac{L_{\delta}}{\xi_{\infty}^{\delta}(\tau)}\right)^{\frac{1}{\nu}},\qquad\delta\in\{\L,\gm,\M\},\label{eq:x_delta_0}
\end{equation}
\eS{}with geometric mean correlation length $\xi_{\infty}^{\gm}$
satisfying $\xi_{\infty}^{\vphantom{\M}\gm\:d}\Def\xi_{\infty}^{\vphantom{\M}\L}\xi_{\infty}^{\M\:d-1}$,
and the system is completely described by the two scaling variables
$x_{\delta}$ and $\rho$ in the FSS limit. The three variables $x_{\delta}$
obey the identity
\begin{equation}
\xL\rho^{\frac{1}{d\nu}-\frac{1}{\nu}}=x_{\gm}=\xM\rho^{\frac{1}{d\nu}}.\label{eq:x-identity}
\end{equation}

Following Fisher \& de Gennes \cite{FisherdeGennes78}, the residual
free energy (\ref{eq:F-res}) then fulfills the scaling \emph{ansatz}
\cite{HuchtGruenebergSchmidt11,HobrechtHucht16a}
\begin{equation}
\Fres(\LL,\LM)\simeq\rho^{1-d}\Theta_{\L}(\xL,\rho)=\Theta_{\gm}(x_{\gm},\rho)=\rho\,\Theta_{\M}(\xM,\rho),\label{eq:F-res-scaling}
\end{equation}
with the universal Casimir potential scaling functions $\Theta_{\delta}$.
Note that both $\Theta_{\L}(\xL,\rho\to0)$ and $\Theta_{\M}(\xM,\rho\to\infty)$
are finite by construction, while $\Theta_{\gm}(x_{\gm},\rho)$ diverges
in both limits \cite{HuchtGruenebergSchmidt11}. 

The Casimir force in $\L$ direction per area $\LM^{d-1}$, 
\begin{equation}
\FC(\LL,\LM)\Def-\frac{1}{\LM^{d-1}}\frac{\partial}{\partial\LL}\Fres(\LL,\LM),\label{eq:FC^L}
\end{equation}
satisfies the finite-size scaling form 
\begin{equation}
\FC(\LL,\LM)\simeq L_{\delta}^{-d}\vartheta_{\delta}(x_{\delta},\rho)\label{eq:vartheta_def}
\end{equation}
for all three cases $\delta\in\{\L,\gm,\M\}$, leading to the identities
\cite{HuchtGruenebergSchmidt11,HobrechtHucht16a}
\begin{equation}
\rho^{1-d}\vartheta_{\L}(\xL,\rho)=\vartheta_{\gm}(x_{\gm},\rho)=\rho\,\vartheta_{\M}(\xM,\rho)\label{eq:vartheta-scalrel}
\end{equation}
analogous to (\ref{eq:F-res-scaling}). We conclude with three scaling
relations between $\Theta_{\delta}$ and $\vartheta_{\delta}$ \cite{HuchtGruenebergSchmidt11,HobrechtHucht16a},
\bS
\begin{align}
\vartheta_{\L}(\xL,\rho) & =-\left[1-d+\frac{1}{\nu}\frac{\xL\partial}{\partial\xL}+\frac{\rho\partial}{\partial\rho}\right]\Theta_{\L}(\xL,\rho)\label{eq:scalrel_perp}\\
\vartheta_{\gm}(x_{\gm},\rho) & =-\left[\frac{1}{d\nu}\frac{x_{\gm}\partial}{\partial x_{\gm}}+\frac{\rho\partial}{\partial\rho}\right]\Theta_{\gm}(x_{\gm},\rho)\label{eq:scalrel_gm}\\
\vartheta_{\M}(\xM,\rho) & =-\left[1+\frac{\rho\partial}{\partial\rho}\right]\Theta_{\M}(\xM,\rho)=-\frac{\partial}{\partial\rho}\big[\rho\,\Theta_{\M}(\xM,\rho)\big].\label{eq:scalrel_para}
\end{align}

\eS{}The FSS functions defined above are universal and only depend
on the bulk and surface universality classes, the system shape and
the boundary conditions. This universality was clearly demonstrated
in \cite{Hucht07a}, where the Casimir force scaling function $\vartheta(x,0)$
of the XY universality class with Dirichlet boundary conditions showed
quantitative agreement between experiments on liquid $^{4}\mathrm{He}$
at the $\lambda$ transition \cite{GarciaChan99} and Monte Carlo
simulations of the classical XY spin model on a simple cubic lattice
\cite{Hucht07a}, both in the thin film limit $\rho\to0$. Subsequent
theoretical studies \cite{VasilyevGambassiMaciolekDietrich07,VasilyevGambassiMaciolekDietrich09}
as well as experiments on binary liquid mixtures \cite{FukutoYanoPershan05,HertleinHeldenGambassiDietrichBechinger08,GamMacHerNelHelBechDiet09}
demonstrated the universal behavior within the framework of the Ising
universality class, for an overview see, e.g., \cite{Gambassi09a}.

\section{Scaling functions for the considered geometry}

In a transfer matrix (TM) formulation as utilized in part I, the system
can have arbitrary real length in propagation direction of the TM,
while the other $d-1$ lengths are fixed. Therefore, we identify $\L$
with the propagation direction and use the scaling variable $\xM$
and the corresponding scaling functions $\Theta_{\M}(\xM,\rho)$ and
$\vartheta_{\M}(\xM,\rho)$ for the description of the FSS behavior.
In the following we will usually drop the index $_{\M}$ from the
quantities $\xM\Def x$, $\Theta_{\M}\Def\Theta$ and $\vartheta_{\M}\Def\vartheta$
for simplicity.

Combining the residual free energy decomposition (\ref{eq:F-res-sum})
with the scaling form (\ref{eq:F-res-scaling}) we now discuss the
according decomposition of the scaling functions $\Theta(x,\rho)$
and $\vartheta(x,\rho)$ for the considered Ising model on the rectangle.
Note that this discussion can be generalized to higher dimensions
within a TM formulation. According to (\ref{eq:F-res-sum}) and (\ref{eq:F-res-scaling})
we decompose $\Theta$ into three parts,
\begin{equation}
\Theta(x,\rho)=\Thetaoo(x)+\rho^{-1}\Theta_{\mathrm{s,c}}(x)+\scalF(x,\rho),\label{eq:Theta-sum}
\end{equation}
where
\begin{equation}
\Thetaoo(x)\simeq\LM f_{\mathrm{b,s}}^{\mathrm{res}}(\LM)\label{eq:f_b,s-res-scal}
\end{equation}
is identified with the known Casimir potential scaling function for
open boundary conditions in strip geometry $\rho\to\infty$ \cite{Au-YangFisher80,EvansStecki94,BrankovDantchevTonchev00},
\begin{equation}
\Thetaoo(x)\Def-\frac{1}{2\pi}\int_{0}^{\infty}\dd\omega\,\log\!\left(1+\frac{\sqrt{x^{2}+\omega^{2}}-x}{\sqrt{x^{2}+\omega^{2}}+x}\ee^{-2\sqrt{x^{2}+\omega^{2}}}\right),\label{eq:Theta-oo}
\end{equation}
which becomes independent from the BCs in $\L$ direction in this
limit and can therefore be calculated from the known exact solution
with periodic BCs in $\L$ direction \cite{HobrechtHucht16a}. The
second term in (\ref{eq:Theta-sum}) contains the surface contributions
from the $\M$ edges as well as the corner contributions, 
\begin{equation}
\Theta_{\mathrm{s,c}}(x)\simeq\Fscres(\LM),\label{eq:Theta_s,c-def}
\end{equation}
is independent of $\rho$ and will be discussed in chapter \ref{subsec:Potential}.
Finally, the third term in (\ref{eq:Theta-sum}) describes the strip
residual free energy contribution, which fulfills
\begin{equation}
\rho\,\scalF(x,\rho)\Def-\log\scalZ(x,\rho)\simeq\Fsres(\LL,\LM),\label{eq:Psi-def}
\end{equation}
where the scaling function $\scalZ$ of the strip residual partition
function \cI{83} is given by
\begin{equation}
\scalZ(x,\rho)\simeq\Zsres(\LL,\LM)=\ee^{-\Fsres(\LL,\LM)}\label{eq:Z-scal}
\end{equation}
 and will be computed in the next chapter. From the limit $\scalF(x,\rho\to\infty)\to0$
\cite{Hucht16a} we get the expected result 
\begin{equation}
\lim_{\rho\to\infty}\Theta(x,\rho)=\Thetaoo(x).\label{eq:Theta-rhoinf}
\end{equation}

According to (\ref{eq:FC-sum}), the total Casimir force scaling function
(\ref{eq:vartheta_def}) can be decomposed to 
\begin{equation}
\vartheta(x,\rho)=-\Thetaoo(x)+\psi(x,\rho),\label{eq:vartheta-sum}
\end{equation}
where the strip Casimir force from (\ref{eq:FC-sum}),
\begin{equation}
\FC_{\strip}(\LL,\LM)\Def-\frac{1}{\LM^{d-1}}\frac{\partial}{\partial\LL}\Fsres(\LL,\LM),\label{eq:FC_strip}
\end{equation}
contributes to the second term,
\begin{equation}
\psi(x,\rho)\Def\frac{\partial}{\partial\rho}\log\scalZ(x,\rho)\simeq\LM^{d}\FC_{\strip}(\LL,\LM),\label{eq:FC_strip-scal}
\end{equation}
while the other contribution 
\begin{align}
\Thetaoo(x)\simeq-\LM^{d}\FC_{\mathrm{b,s}}(\LM) & =\LM\frac{\partial}{\partial\LL}[\LL\fbsres(\LM)]=\LM\fbsres(\LM)\label{eq:FC_b,s-scal}
\end{align}
is again known (\ref{eq:Theta-oo}) in our case. Note that the total
Casimir force scaling function (\ref{eq:vartheta-sum}) could also
have been obtained using the scaling relation (\ref{eq:scalrel_para}).
In the following, we will first calculate the FSS function $\scalZ(x,\rho)$
and determine the Casimir force scaling function $\vartheta(x,\rho)$. 

\section{Finite-size scaling limit of the strip residual partition function}

\subsection{The characteristic polynomial}

\begin{figure}
\begin{centering}
\includegraphics{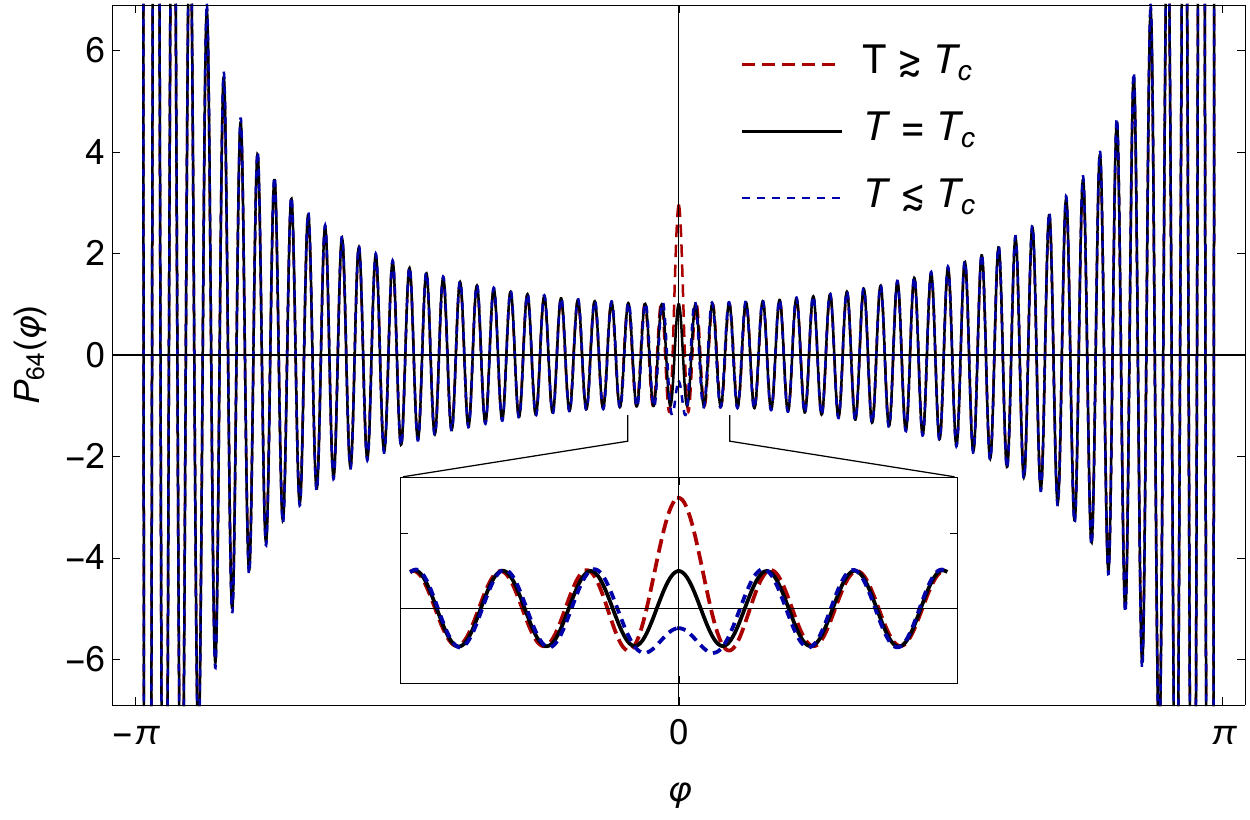}
\par\end{centering}
\caption{Characteristic polynomial $P_{M}(\varphi)$, Eq.~\cI{45}, for $M=64$
and three different temperatures above, at, and below $\protect\Tc$.
The polynomials differ only around $\varphi=0$, see inset. \label{fig:CP-32}}
\end{figure}
We now turn back to the $2d$ Ising model on the $L\times M$ rectangle
discussed in I and revert the variables $\LL$ and $\LM$ back to
$L\Def\LL$ and $M\Def\LM$. We are allowed to assume isotropic couplings
$t^{*}=z$ in the FSS limit in order to keep things simple, as near
criticality a coupling anisotropy $K^{\L}\neq K^{\M}$ can be compensated
by a scale transformation of the real variable $L\mapsto L\xi_{+}^{\M}/\xi_{+}^{\L}$
in conjunction with $K^{\L}\mapsto K^{\M}$, without changing the
generalized aspect ratio $\rho$ \cite{Hucht02a}. Then, the critical
point is at $z=\zc\Def\sqrt{2}-1$, and we can use $\tau=1-z/\zc$
for the reduced temperature, with corresponding isotropic correlation
length amplitude $\xi_{+}=1/2$. The resulting FSS variables $x$
and $\rho$ from (\ref{eq:x,rho_0}) in terms of the relevant length
$M$ become
\begin{equation}
x=2M\left(1-\frac{z}{\zc}\right),\qquad\rho=\frac{L}{M}.\label{eq:x,rho}
\end{equation}
We now perform the FSS limit of the results from I by replacing all
quantities with the $M$-dependent scaling forms and then performing
the limit $M\to\infty$ with fixed $x$ and $\rho$. As the strip
residual partition function $\Zsres$ is convergent in this limit,
we do not encounter regularization problems as in the case of the
edge contribution $\Fsc$ discussed later. We will use the tilde $\scal{\cdot}$
for quantities in the FSS limit and capitals for scaling variables.

\begin{figure}
\begin{centering}
\includegraphics{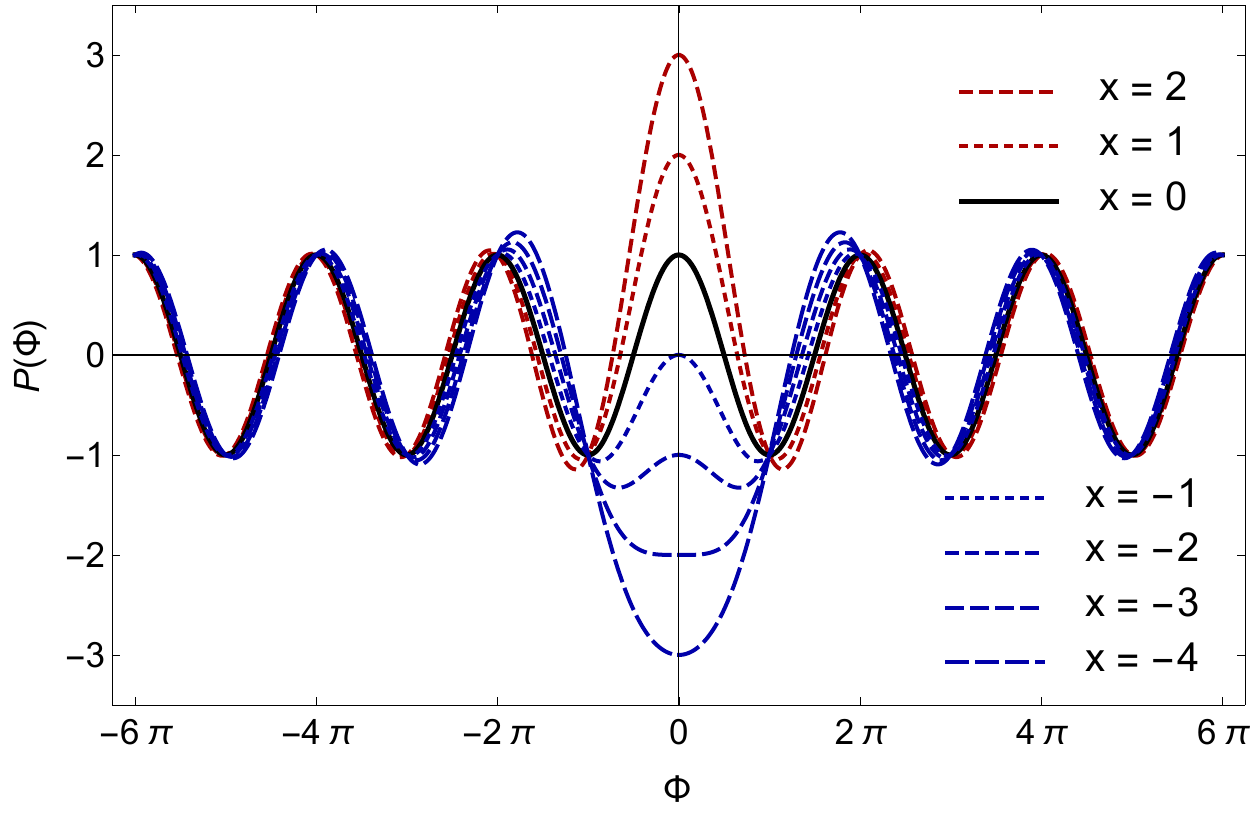}
\par\end{centering}
\caption{Universal characteristic polynomial $P(\Phi)$, Eq.~(\ref{eq:P(Phi)}),
for different scaled temperatures $x$ in the finite-size scaling
limit. The first zero is doubly degenerate at $x=-1$ and becomes
imaginary below. \label{fig:CP}}
\end{figure}

Therefore, we replace the temperature variable $z$ and the length
$L$ according to 
\begin{equation}
z\mapsto\zc\left(1-\frac{x}{2M}\right),\qquad L\mapsto\rho M,\label{eq:resc-z-L}
\end{equation}
and first turn to the characteristic polynomial $P_{M}(\varphi)$
\cI{45}. For large $M$ the relevant scaling contributions are obtained
by a rescaling of the angle variable $\dom{\varphi}$ \cI{86} according
to
\begin{equation}
\dom{\varphi}\mapsto\frac{\Phi}{M},\label{eq:resc-phi}
\end{equation}
which immediately leads to the remarkably simple universal FSS form
of the characteristic polynomial \cI{45},
\begin{equation}
P(\Phi)\Def\cos\Phi+\frac{x}{\Phi}\sin\Phi,\label{eq:P(Phi)}
\end{equation}
with infinitely many zeroes $\Phi_{\mu}$, $\mu\in\mathbb{N}$. All
zeroes $\Phi_{\mu}$ are real and positive except $\Phi_{1}$, which
is zero for $x=-1$ and becomes imaginary for $x<-1$, see figure~\ref{fig:CP}
and table \ref{tab:zeroes}. 

Comparing $P(\Phi)$ with $P_{M}(\varphi)$ of a finite system as
shown in figure~\ref{fig:CP-32}, we notice the following: while
for small $|\varphi|\lesssim\pi/2$ both curves become more and more
similar for large $M$, the deviations for $|\varphi|>\pi/2$ stem
from the lattice dispersion encoded in $P_{M}(\varphi)$, which is
different from the continuum dispersion of $P(\Phi)$. We will see
below that a careful regularization of the resulting infinite products
is necessary in order to overcome the UV singularities emerging in
the FSS limit.

\begin{table}
\setlength\tabcolsep{3ex}
\begin{tabular}{c|c|c|c|c}
$x$ & $\Phi_1$ & $\Phi_2$ & $\Phi_3$ & $\Phi_4$ \\
\hline
$-4$ & 3.997302692 i & 3.916435368 & 7.355927023 & 10.63585142 \\
$-3$ & 2.984704585 i & 4.078149765 & 7.472192660 & 10.72277106 \\
$-2$ & 1.915008048 i & 4.274782271 & 7.596546020 & 10.81267333 \\
$-1$ & 0             & 4.493409458 & 7.725251837 & 10.90412166 \\
$ 0$ & $\pi/2$       & $3\pi/2$    & $5\pi/2$    & $7\pi/2$    \\
$ 1$ & 2.028757838   & 4.913180439 & 7.978665712 & 11.08553841 \\
$ 2$ & 2.288929728   & 5.086985094 & 8.096163603 & 11.17270587 \\
$ 3$ & 2.455643863   & 5.232938454 & 8.204531363 & 11.25604301 \\
$ 4$ & 2.570431560   & 5.354031841 & 8.302929183 & 11.33482558 \\
\end{tabular}

\caption{Location of the first few zeroes of $P(\Phi)$, (\ref{eq:P(Phi)}),
for $\mu=1,\ldots,4$ and several values of $x$. \label{tab:zeroes}}
\end{table}
The scaling limit of the eigenvalues $\dom{\lambda}$, expressed through
the Onsager-$\dom{\gamma}$, is determined from the isotropic version
of \cI{43},
\begin{equation}
\cos\dom{\varphi}=z_{+}^{*}z_{+}^{\vphantom{*}}-\cosh\dom{\gamma},\label{eq:cos_phi_iso}
\end{equation}
under the rescaling
\begin{equation}
\dom{\gamma}\mapsto\frac{\Gamma}{M},\qquad\dom{\lambda}^{-L}=\ee^{-L\dom{\gamma}}\mapsto\ee^{-\rho\Gamma},\label{eq:sc-gamma}
\end{equation}
to be 
\begin{equation}
\Gamma=\sqrt{x^{2}+\Phi^{2}}.\label{eq:Gamma}
\end{equation}
At the zeroes $\Phi_{\mu}$ we find the simple relation
\begin{equation}
\Phi_{\mu}\cot\Phi_{\mu}=\sigma_{\mu}\Gamma_{\mu}\cos\Phi_{\mu}=-x,\label{eq:Gamma_cos_Phi}
\end{equation}
with parity \cI{76}
\begin{equation}
\sigma_{\mu}\Def(-1)^{\mu-1}.\label{eq:parity}
\end{equation}
The two remaining quantities entering the residual matrix $\mat Y$
\cI{87b}, $p_{\mu}$ and $v_{\mu}$ from \cI{74} and \cI{87a},
are calculated in the next section.

\subsection{Contour integration}

\begin{figure}
\begin{centering}
\includegraphics[width=0.9\columnwidth]{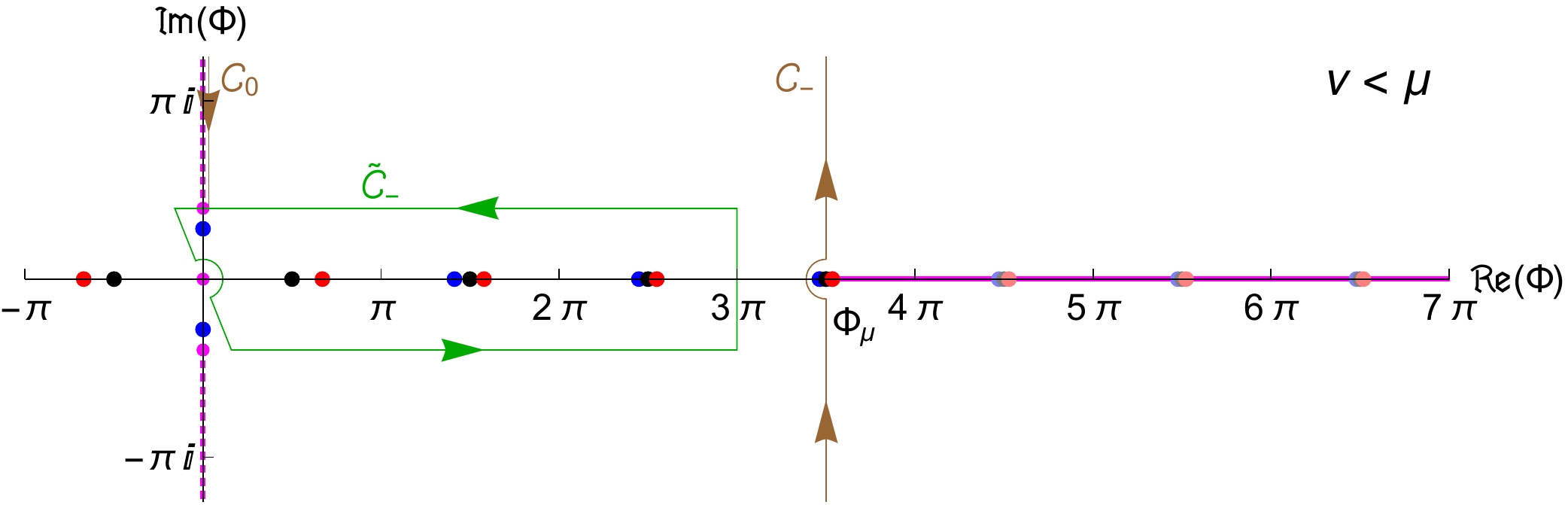}
\par\end{centering}
\begin{centering}
\includegraphics[width=0.9\columnwidth]{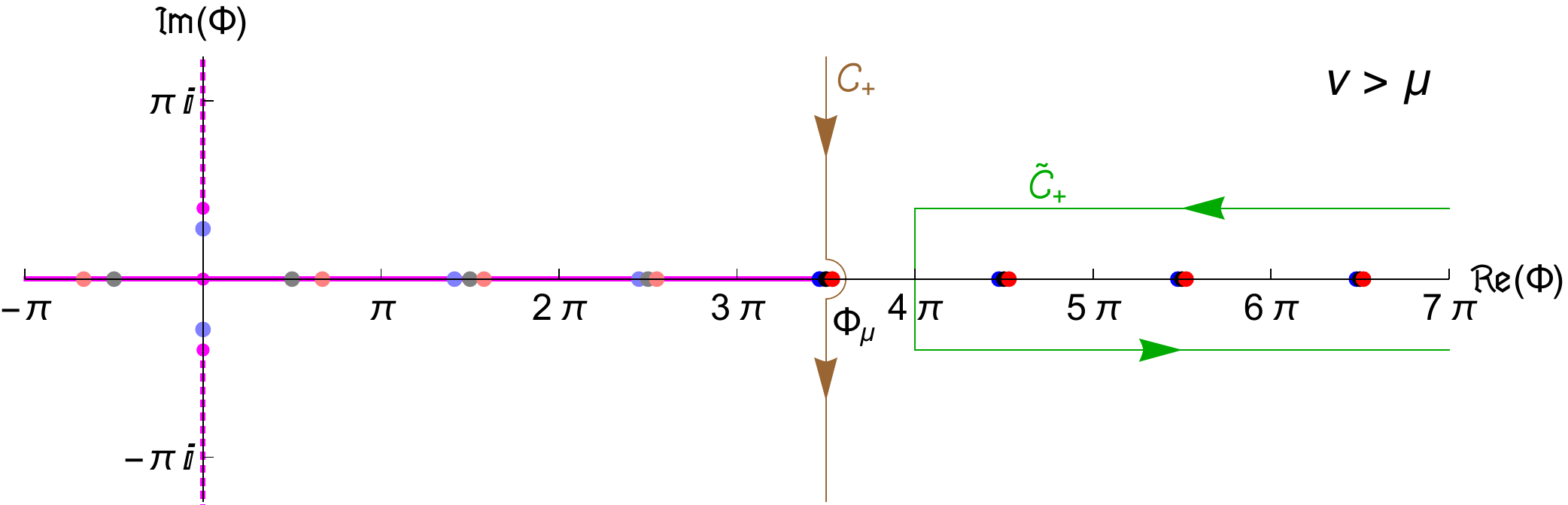}
\par\end{centering}
\caption{Complex zeroes of $P(\Phi)$ for three temperatures $x=\{-\frac{5}{4},0,\frac{5}{4}\}$
below (blue), at (black), and above (red) the critical point $\protect\Tc$,
for $\mu=4$. At $x=-1$ the first zero $\Phi_{1}=0$. The upper/lower
plot shows the integration contours $\tilde{C}_{\pm}$ from (\ref{eq:sc-p_mu_int})
(green lines) as well as $C_{0}$ from (\ref{eq:sc-p_mu_2}) and $C_{\pm}$
from (\ref{eq:int_id_1}) (brown lines) for the terms $\nu\lessgtr\mu$,
respectively. The log/sqrt branch cuts are shown as solid/dashed magenta
lines. \label{fig:contours}}
\end{figure}
We first calculate the regularized scaling limit $\scal p_{\mu}$
of the product $p_{\mu}$ \cI{74} using the rescaling 
\begin{equation}
c_{\mu}-c_{\nu}=\cos\dom{\varphi}_{\mu}-\cos\dom{\varphi}_{\nu}\mapsto\cos\frac{\Phi_{\mu}}{M}-\cos\frac{\Phi_{\nu}}{M}=\frac{\Phi_{\nu}^{2}-\Phi_{\mu}^{2}}{2M^{2}}+\mathcal{O}(M^{-4}).\label{eq:c_mu-c_nu}
\end{equation}
An important simplification stems from the fact that the residual
matrix $\mat Y$ only contains products $p_{\mu\in\So}p_{\mu'\in\Se}$
of one odd and one even factor, respectively, such that $\mu$-independent
terms in $\left(\cdot\right)^{-\sigma_{\mu}\sigma_{\nu}}$ cancel
in the resulting products. We can therefore drop the factor $1/2M^{2}$
and instead insert a regularizing denominator, that ensures the convergence
of the infinite product, to get 
\begin{equation}
\scal p_{\mu}\Def\lim_{N\to\infty}\frac{{\prod'}_{\nu=1}^{N}\left(\Phi_{\nu}^{2}-\Phi_{\mu}^{2}\right)^{-\sigma_{\mu}\sigma_{\nu}}}{\prod_{\nu=1}^{N}\left(\Phi_{\nu}^{2}\right)^{-\sigma_{\mu}\sigma_{\nu}}}=\Phi_{\mu}^{2}\prod_{\nu=1}^{\infty}{\vphantom{\prod}}'\left(1-\frac{\Phi_{\mu}^{2}}{\Phi_{\nu}^{2}}\right)^{-\sigma_{\mu}\sigma_{\nu}}.\label{eq:sc-p_mu_def}
\end{equation}
Here, $\prod'$ denotes the regularized product, with zero and infinite
factors removed. This alternating product over the zeroes $\Phi_{\nu}$
of $P(\Phi)$ (\ref{eq:P(Phi)}) can be calculated by complex contour
integration using Cauchy's residue theorem: we first rewrite the product
as two sums, split into terms $\nu{<}\mu$ and $\nu{>}\mu$, respectively,
as we have to avoid the zero $\Phi_{\mu}$. In the first sum, the
argument of the $\log$ is negated to always be positive, leading
to an extra overall factor $\sigma_{\mu}$. This circumvents the cut
of the logarithm in the complex plane (solid magenta lines on the
real axes in figure~\ref{fig:contours}) in the resulting contour
integrals. We find \bS
\begin{align}
\scal p_{\mu} & =\Phi_{\mu}^{2}\,\exp\!\left[-\sigma_{\mu}\sum_{\nu=1}^{\infty}{\vphantom{\sum}}'\sigma_{\nu}\log\!\left({\textstyle 1-\frac{\Phi_{\mu}^{2}}{\Phi_{\nu}^{2}}}\right)\right]\\
 & =\sigma_{\mu}\Phi_{\mu}^{2}\exp\!\text{\ensuremath{\left[-\sigma_{\mu}\left\{  \sum_{\nu=1}^{\mu-1}\sigma_{\nu}\log\!\left({\textstyle \frac{\Phi_{\mu}^{2}}{\Phi_{\nu}^{2}}-1}\right)+\sum_{\nu=\mu+1}^{\infty}\sigma_{\nu}\log\!\left({\textstyle 1-\frac{\Phi_{\mu}^{2}}{\Phi_{\nu}^{2}}}\right)\right\}  \right]}}\\
 & =\sigma_{\mu}\Phi_{\mu}^{2}\exp\!\text{\ensuremath{\left[-\sigma_{\mu}\sum_{\pm}\frac{1}{2\pi\ii}\oint_{\tilde{C}_{\pm}}\dd\Phi\log\!\left(\pm{\textstyle \frac{\Phi^{2}-\Phi_{\mu}^{2}}{\Phi^{2}}}\right)R(\Phi) \right]}},\label{eq:sc-p_mu_int}
\end{align}
\eS{}with alternating counting polynomial $R(\Phi)$ fulfilling 
\begin{equation}
\Res{\Phi=\Phi_{\nu}}{R(\Phi)}=\sigma_{\nu},\label{eq:Res-condition}
\end{equation}
which will be constructed in the next chapter. The two contours $\tilde{C}_{\pm}$
enclose the respective zeroes $\Phi_{\nu}$ and are shown as green
lines in figure~\ref{fig:contours}.

\subsection{Construction of counting polynomial $R(\Phi)$}

The alternating counting polynomial $R(\Phi)$ satisfying (\ref{eq:Res-condition})
is constructed in the following way: we discriminate the even and
odd zeroes by first rewriting\footnote{The case $\mu=1$, $x<-1$, with imaginary $\Phi_{1}$, has to be
handled separately, but leads to the same result for $P_{\pm}$.} 
\begin{equation}
P(\Phi)=\cos\Phi+\frac{x}{\Phi}\sin\Phi=\Re\left[\ee^{\ii\Phi}\left(1+\frac{x}{\ii\Phi}\right)\right].\label{eq:P(Phi)_complex}
\end{equation}
Normalizing the modulus of $1+x/(\ii\Phi)$ to one, the zeroes of
the real part move to $\pm\ii$, and the condition for the odd $(+)$
and even $(-)$ zeroes becomes
\begin{equation}
P_{\pm}(\Phi)\Def1\pm\ee^{\ii\Phi}\Gamma^{-1}(x+\ii\Phi)\stackrel{!}{=}0,\label{eq:P_pm}
\end{equation}
with $\Gamma=\sqrt{x^{2}+\Phi^{2}}$ from (\ref{eq:Gamma}), leading
to the odd and even counting polynomials
\begin{equation}
R_{\pm}(\Phi)\Def Q_{\pm}'(\Phi),\label{eq:R_pm}
\end{equation}
fulfilling $\Res{\Phi=\Phi_{\nu}}{R_{\pm}(\Phi)}=\delta_{\sigma_{\nu},\pm1}$,
with symmetrized antiderivatives
\begin{equation}
Q_{\pm}(\Phi)\Def\frac{1}{2}\log[P_{\pm}(\Phi)P_{\pm}(-\Phi)]=\frac{1}{2}\log\left[2\left(1\pm\frac{x}{\Gamma}\cos\Phi\mp\frac{\Phi}{\Gamma}\sin\Phi\right)\right].\label{eq:Q_pm}
\end{equation}
For the alternating counting polynomial $R(\Phi)\Def Q'(\Phi)$ we
find the antiderivatives 
\begin{equation}
Q(\Phi)\Def Q_{+}(\Phi)-Q_{-}(\Phi)=\artanh\left(\frac{x}{\Gamma}\cos\Phi-\frac{\Phi}{\Gamma}\sin\Phi\right),\label{eq:Q_def}
\end{equation}
 leading to the result 
\begin{equation}
R(\Phi)=-\sqrt{\frac{\Gamma^{2}}{\Phi^{2}}}\frac{1+x\Gamma^{-2}}{P(\Phi)}\stackrel{\Phi\in\mathbb{C}^{+}}{=}-\frac{\Gamma}{\Phi}\frac{1+x\Gamma^{-2}}{P(\Phi)}=\frac{-(x+x^{2}+\Phi^{2})}{\sqrt{x^{2}+\Phi^{2}}(\Phi\cos\Phi+x\sin\Phi)},\label{eq:R}
\end{equation}
where $\mathbb{C}^{+}$ denotes the complex domain $\{z\in\mathbb{C}\,|\,-\frac{\pi}{2}<\arg z\leq\frac{\pi}{2}\vee z=0\}$.
The resulting analytic structure of the integrands is shown in figure~\ref{fig:contours}.
$R(\Phi)$ has square root branch cuts from $\pm\ii x$ to $\pm\ii\infty$
as well as a simple pole at zero, with residuum $\Res{\Phi=0}{R(\Phi)}=-1$.
Additionally, the logarithm contributes log branch cuts running either
from $-\Phi_{\mu}$ to $\Phi_{\mu}$ for $\mathcal{C}_{+}$, or from
$\pm\Phi_{\mu}$ to $\pm\infty$ for $\mathcal{C}_{-}$.

\subsection{Calculation of the integrals}

We now deform the contour $\tilde{C}_{-}$ to $C_{0}$ at the imaginary
axis and to the line $C_{-}$ and move $\tilde{C}_{+}$ to $C_{+}$
as shown in figure \ref{fig:contours}. The resulting contribution
from $C_{\pm}$ can be calculated using (\ref{eq:Q_def}) and reads
\begin{equation}
-\sigma_{\mu}\sum_{\pm}\frac{1}{2\pi\ii}\int_{\mathcal{C}_{\pm}}\dd\Phi\log\!\left(\pm\frac{\Phi^{2}-\Phi_{\mu}^{2}}{\Phi^{2}}\right)R(\Phi)=-\log\!\left[\frac{\Phi_{\mu}}{4}\left(1+x\Gamma_{\mu}^{-2}\right)\right].\label{eq:int_id_1}
\end{equation}
The pole at zero and the special behavior of $\Phi_{1}$ at $x=-1$
has to be carefully analyzed, leading to the result
\begin{equation}
\scal p_{\mu}=\frac{4\sigma_{\mu}\Phi_{\mu}}{1+x\Gamma_{\mu}^{-2}}\left(\frac{x+1}{2x}\Phi_{\mu}\right)^{-\sigma_{\mu}}\exp\!\left[\frac{\sigma_{\mu}}{\pi\ii}\int_{\ii|x|}^{\ii\infty}\dd\Phi\log\!\left(\left(\frac{x+1}{2x}\right)^{-2}\frac{\Phi_{\mu}^{2}-\Phi^{2}}{\Phi^{2}\Phi_{\mu}^{2}}\right)R(\Phi)\right],\label{eq:sc-p_mu_2}
\end{equation}
where we have used the identity, c.\,f. (\ref{eq:Q_def}),
\begin{equation}
\frac{1}{\pi\ii}\int_{\ii|x|}^{\ii\infty}\dd\Phi R(\Phi)=\frac{1}{2}(\sign x-1).\label{eq:int_id_2}
\end{equation}
In (\ref{eq:sc-p_mu_2}) we can again drop ($\mu$-independent terms)$^{\sigma_{\mu}}$
to get
\begin{equation}
\scal p_{\mu}^{\dagger}\Def\frac{4\sigma_{\mu}\Phi_{\mu}^{1-\sigma_{\mu}}}{1+x\Gamma_{\mu}^{-2}}\exp\!\left[\frac{\sigma_{\mu}}{\pi\ii}\int_{\ii|x|}^{\ii\infty}\dd\Phi\log\!\left(1-\frac{\Phi^{2}}{\Phi_{\mu}^{2}}\right)R(\Phi)\right].\label{eq:sc-p_mu^dagger}
\end{equation}
We can combine this result for $\scal p_{\mu}^{\dagger}$ with the
scaling form of the coefficients $\dom v_{\mu}$ from \cI{87a}, 
\begin{equation}
\dom v_{\mu}\mapsto\scal v_{\mu}\Def\scal p_{\mu}^{\dagger}\sigma_{\mu}(\Gamma_{\mu}-\sigma_{\mu}x)^{\sigma_{\mu}}\label{eq:sc-v_mu_def}
\end{equation}
to find, as $(\Gamma-x)(\Gamma+x)=\Phi^{2}$, the resulting scaling
form of the matrix elements \bS 
\begin{equation}
\scal v_{\mu}=4\frac{\Gamma_{\mu}-x}{1+x\Gamma_{\mu}^{-2}}\exp\!\left[\frac{\sigma_{\mu}}{\pi\ii}\int_{\ii|x|}^{\ii\infty}\dd\Phi\log\!\left(1-\frac{\Phi^{2}}{\Phi_{\mu}^{2}}\right)R(\Phi)\right].\label{eq:sc-v_mu}
\end{equation}
In the special case $\mu=1$ at $x\to-1$, where $\Phi_{1}\to0$,
the integral diverges logarithmically while the prefactor $\Gamma_{\mu}-x$
goes to zero. Using the series expansion around $x=-1$, $\Phi_{1}=\sqrt{3(x+1)}+\mathcal{O}(x+1)^{3/2}$
we can nonetheless proceed and find
\begin{equation}
\scal v_{1}|_{x=-1}=12\exp\!\left[\frac{1}{\pi\ii}\int_{\ii}^{\ii\infty}\dd\Phi\log\!\left(-\Phi^{2}\right)R(\Phi)\right]=6.39303337215\ldots\,.\label{eq:sc-v_mu-1}
\end{equation}

\eS{}The scaling form of the Cauchy matrix $\mat T$ from \cI{70a}
reads 
\begin{equation}
(\scal{\mat T})_{\mu\nu}\Def\frac{1}{\Phi_{\nu}^{2}-\Phi_{\mu}^{2}}=\frac{1}{\Gamma_{\nu}^{2}-\Gamma_{\mu}^{2}},\label{eq:sc-T}
\end{equation}
where we have moved the factor $2M^{2}$ from the expansion around
$M=\infty$ into the Cauchy determinant \cI{80}. Combining (\ref{eq:sc-T})
with the diagonal matrices 
\begin{equation}
(\mat{\Gamma})_{\mu\mu}\Def\Gamma_{\mu},\qquad(\scal{\mat V})_{\mu\mu}\Def\scal v_{\mu},\label{eq:mat_Gamma_and_V}
\end{equation}
we find the result for the residual matrix $\mat Y$ in the finite-size
scaling limit
\begin{equation}
\scal{\mat Y}(x,\rho)=-\ee^{-\rho\mat{\Gamma}_{\Se}}\scal{\mat V}_{\Se}\scal{\mat T}_{\Se,\So}\ee^{-\rho\mat{\Gamma}_{\So}}\scal{\mat V}_{\So}\scal{\mat T}_{\So,\Se},\label{eq:sc-Y}
\end{equation}
from which we can calculate the universal partition function scaling
function $\scalZ$ and the Casimir potential scaling function $\scalF$
according to \bS
\begin{align}
\scalZ(x,\rho) & =\det[\mat 1+\scal{\mat Y}(x,\rho)],\label{eq:Sigma2}\\
\scalF(x,\rho) & =-\rho^{-1}\log\scalZ(x,\rho).\label{eq:Psi2}
\end{align}
\eS{}Note that $\scalZ$ depends on the aspect ratio $\rho$ only
via the two exponentials in $\scal{\mat Y}(x,\rho)$.

\subsection{Representation of the partition function scaling function $\protect\scalZ(x,\rho)$}

\global\long\def\Sos{\boldsymbol{\mathcal{S}}}
While the scaled residual matrix $\scal{\mat Y}$ is infinite dimensional,
its matrix elements become exponentially small for large $\mu,\nu$
at least if $\rho\gtrsim1$, 
\begin{equation}
(\scal{\mat Y})_{\nu\in\Se,\mu\in\So}=\mathcal{O}(\ee^{-\rho(\Gamma_{1}+\Gamma_{\nu})}),\label{eq:sc-Y_asymp}
\end{equation}
such that we can take the upper left $N\times N$ submatrix for a
rapidly converging calculation of the determinant. This direct approach
is however only applicable for $N\lesssim10$ if $\rho$ is left as
a free parameter, as the symbolic evaluation of a general $N\times N$
determinant is exponentially hard and requires $N!$ terms. However,
we alternatively can expand the determinant according to \cI{97}
and directly calculate all terms $\mathcal{O}(\ee^{-2\pi\rho n})$
up to $n\leq N$. Therefore we define the set $\Sos_{n}$ of all subsets
$\S$ of the natural numbers with equal number of even and odd elements
$\mu$ fulfilling the condition
\begin{equation}
\Sos_{n}=\Big\{\S\,\Big|\,\S\subset\mathbb{N}\,\wedge\,\sum_{\mu\in\S}\sigma_{\mu}=0\,\wedge\,\sum_{\mu\in\S}\left(\mu-{\textstyle \frac{1}{2}}\right)=2n\Big\},\label{eq:Sos}
\end{equation}
 e.g., $\Sos_{1}=\{\{1,2\}\}$ and $\Sos_{4}=\{\{1,8\},\{3,6\},\{5,4\},\{7,2\},\{1,3,2,4\}\}$.
We then can define the $N$-th approximant to the partition function
scaling function \bS\label{eq:Sigma-N}
\begin{equation}
\scalZ^{(N)}(x,\rho)\Def1+\sum_{n=1}^{N}\sum_{\S\in\Sos_{n}}a_{\S}\ee^{-\rho\,\Gamma_{\S}},\label{eq:Sigma-N1}
\end{equation}
with 
\begin{equation}
a_{\S}\Def\prod_{\{\mu,\nu\}\subset\S}\left(\Phi_{\mu}^{2}-\Phi_{\nu}^{2}\right)^{-2\sigma_{\mu}\sigma_{\nu}}\prod_{\mu\in\S}\scal v_{\mu},\qquad\Gamma_{\S}\Def\Tr\mat{\Gamma}_{\S}=\sum_{\mu\in\S}\Gamma_{\mu},\label{eq:Sigma-N-aG}
\end{equation}
\eS{}to get an exponentially precise approximation to the scaling
function (\ref{eq:Sigma2})
\begin{equation}
\scalZ(x,\rho)=\scalZ^{(N)}(x,\rho)+\mathrm{o}\!\left(\ee^{-2\pi\rho N}\right).\label{eq:Z-eq-ZN}
\end{equation}
Note that (\ref{eq:Sigma-N1}) is a perturbative series, and the number
of elements in $\Sos_{n}$ equals the famous integer partition function
$P_{n}$ from number theory \cite{Hardy79}, $|\Sos_{n}|=P_{n}$.
Consequently, the calculation of, e.g., $\scalZ^{(30)}$ requires
only 28629 terms instead of the $30!=2.65\times10^{32}$ terms needed
for the direct evaluation of the determinant.

\begin{table}
\setlength\tabcolsep{8pt}

\begin{tabular}{c|c|cc|cc}
\multirow{2}{*}{order $n$} & \multirow{2}{*}{set $\S$} & \multicolumn{2}{c|}{$x=-1$} & \multicolumn{2}{c}{$x=1$}\tabularnewline
 &  & $a_{\S}$ & $\Gamma_{\S}/2\pi$ & $a_{\S}$ & $\Gamma_{\S}/2\pi$\tabularnewline
\hline 
1 & $\{1,2\}$ & 0.41416034599 & 0.89179907560 & 0.15689480307 & 1.15797017264\tabularnewline
\hline 
2 & $\{3,2\}$ & 0.58023590813 & 1.97241431063 & 0.27677728168 & 2.07776833638\tabularnewline
2 & $\{1,4\}$ & 0.02228040130 & 1.90188245064 & 0.01146254079 & 2.13146302530\tabularnewline
\hline 
3 & $\{3,4\}$ & 0.48130844027 & 2.98249768567 & 0.31674195444 & 3.05126118904\tabularnewline
3 & $\{5,2\}$ & 0.05012321797 & 2.97699865033 & 0.02613926677 & 3.06476730850\tabularnewline
3 & $\{1,6\}$ & 0.00537233691 & 2.90454040035 & 0.00297985504 & 3.12373747328\tabularnewline
\hline 
4 & $\{5,4\}$ & 0.47345042883 & 3.98708202537 & 0.34034540402 & 4.03826016115\tabularnewline
4 & $\{3,6\}$ & 0.04512462939 & 3.98515563538 & 0.03206462405 & 4.04353563702\tabularnewline
4 & $\{7,2\}$ & 0.01444309494 & 3.97874169683 & 0.00782432208 & 4.05964386240\tabularnewline
4 & $\{1,8\}$ & 0.00206454953 & 3.90577401397 & 0.00118447481 & 4.12008924457\tabularnewline
4 & $\{1,3,2,4\}$ & 0.31379568621 & 3.87429676127 & 0.07798039866 & 4.20923136168\tabularnewline
\end{tabular}

\caption{Amplitudes $a_{\protect\S}$ and exponents $\Gamma_{\protect\S}$
from (\ref{eq:Sigma-N1}) for $x=\pm1$ and $n=1,\ldots,4$. \label{tab:amplitudes_exponents}}
\end{table}
In table \ref{tab:amplitudes_exponents} the leading coefficients
$a_{\S}$ and $\Gamma_{\S}$ are given for the two cases $x=\pm1$.
Already with these few terms the error is smaller than $\ee^{-8\pi}\approx10^{-11}$
for all $\rho\geq1$, while the case $\rho<1$ can be calculated using
the symmetry under exchange of the two directions $\M$ and $\L$
\cite[(52)]{HuchtGruenebergSchmidt11}, 
\begin{equation}
\Theta(x,\rho)=\rho^{-2}\Theta(x\rho,\rho^{-1}).\label{eq:Theta_symmetry}
\end{equation}
With these expressions we now present results for the Casimir scaling
functions. We first turn to the critical point $x=0$.

\section{Results }

\subsection{Results at $x=0$}

\global\long\def\zero{^{(0)}}
\global\long\def\Beta{\mathrm{B}}
At criticality $x=0$ we work with the volume FSS functions $\Sigma_{\gm}$,
$\Psi_{\gm}$ and $\psi_{\gm}$ in order to compare with conformal
field theory (CFT) results. Remember that $\scalZ(0,\rho)=\scalZ_{\gm}(0,\rho)$,
$\rho\,\Psi(0,\rho)=\Psi_{\gm}(0,\rho)$ and $\psi(0,\rho)=\psi_{\gm}(0,\rho)$.
We use the superscript $\zero$ for quantities at $x=0$. For $x=0$
the characteristic polynomial reduces to 
\begin{equation}
P\zero(\Phi)=\cos\Phi,\label{eq:P0}
\end{equation}
with trivial zeroes $\Phi_{\mu}\zero=(\mu-\frac{1}{2})\pi,\;\mu\in\mathbb{N}$.
Consequently, the infinite product $\scal p_{\mu}^{\dagger}$ from
(\ref{eq:sc-p_mu^dagger}) can be calculated exactly at $x=0$ and
can be expressed through the Euler beta function $\Beta(a,b)=\Gamma(a)\Gamma(b)/\Gamma(a+b)$,
with the result 
\begin{equation}
\scal p_{\mu}^{\dagger\,(0)}=4\sigma_{\mu}\Phi_{\mu}\zero\left[\frac{1}{\sqrt{2}\pi}\Beta\!\left(\frac{\mu}{2},\frac{1}{2}\right)\right]^{2\sigma_{\mu}}.\label{eq:p_mu^dagger_0}
\end{equation}
The resulting coefficients $\scal v_{\mu}$ (\ref{eq:sc-v_mu}) become
\begin{equation}
\scal v_{\mu}\zero=\scal p_{\mu}^{\dagger\,(0)}\sigma_{\mu}\left(\Phi_{\mu}\zero\right)^{\sigma_{\mu}}=4\left(\Phi_{\mu}\zero\right)^{1+\sigma_{\mu}}\left[\frac{1}{\sqrt{2}\pi}\Beta\!\left(\frac{\mu}{2},\frac{1}{2}\right)\right]^{2\sigma_{\mu}},\label{eq:v_mu_0}
\end{equation}
and with 
\begin{equation}
\scal{\mat Y}(0,\rho)=-\ee^{-\rho\mat{\Phi}_{\Se}\zero}\scal{\mat V}_{\Se}\zero\scal{\mat T}_{\Se,\So}^{(0)}\ee^{-\rho\mat{\Phi}_{\So}\zero}\scal{\mat V}_{\So}\zero\scal{\mat T}_{\So,\Se}^{(0)}\label{eq:tildeY(0)}
\end{equation}
from (\ref{eq:sc-Y}) and (\ref{eq:Sigma2}) we arrive at the rapidly
converging series 
\begin{equation}
\scalZ_{\gm}(0,\rho)=1+\frac{1}{4}e^{-2\pi\rho}+\frac{13}{32}e^{-4\pi\rho}+\frac{55}{128}e^{-6\pi\rho}+\frac{1235}{2048}e^{-8\pi\rho}+\frac{4615}{8192}e^{-10\pi\rho}+\mathcal{O}(e^{-12\pi\rho}).\label{eq:sc-Z_series}
\end{equation}
This result is identical to the prediction from conformal field theory
\cite{KlebanVassileva91}, which can be written in several ways, 
\begin{equation}
\scalZ_{\gm}(0,\rho)=e^{-\frac{\pi}{48}\rho}\eta^{-\frac{1}{4}}(\ii\rho)=(e^{-2\pi\rho})_{\infty}^{-\frac{1}{4}}=\QP\left(\left.-\tfrac{1}{4}\right|e^{-2\pi\rho}\right),\label{eq:Sigma(0)}
\end{equation}
with the Dedekind eta function $\eta$, the $q$-Pochhammer symbol
$(q)_{\infty}$, or in terms of the $q$-products introduced in \cI{A2}.
Here we have taken into account the additional contribution $e^{-\frac{\pi}{48}\rho}$
from the strip geometry, see below for details. While we were not
able to proof this correspondence, the series coefficients agree at
least for the first 30 terms and we have no doubt about the equivalency.

Observing the fact that for both even and odd $\mu$, $\scal v_{\mu}\zero$
is $2\pi^{2}$ times a squared rational number, we define the square
root \bS
\begin{align}
\scal w_{\mu}^{(0)}\Def\sqrt{\scal v_{\mu}\zero} & =2\left(\Phi_{\mu}\zero\right)^{\frac{1+\sigma_{\mu}}{2}}\left[\frac{1}{\sqrt{2}\pi}\Beta\!\left(\frac{\mu}{2},\frac{1}{2}\right)\right]^{\sigma_{\mu}}\label{eq:tildew1}\\
 & =\sqrt{2}\sigma_{\mu}\left(\mu-{\textstyle \frac{1}{2}}\right)^{\frac{1+\sigma_{\mu}}{2}}\Beta\!\left(\frac{1}{2}\left[\mu+\frac{1-\sigma_{\mu}}{2}\right],\frac{\sigma_{\mu}}{2}\right)\label{eq:tildew2}
\end{align}
\eS{}as $\Beta\!\left({\textstyle \frac{\mu}{2}},{\textstyle \frac{1}{2}}\right)^{-1}=-\frac{1}{2\pi}\Beta\!\left({\textstyle \frac{\mu+1}{2}},-{\textstyle \frac{1}{2}}\right).$
The generating function $\mathcal{W}^{(0)}(\eta)$ of the coefficients
$\scal w_{\mu}^{(0)}$ can also be given and reads
\begin{equation}
\mathcal{W}^{(0)}(\eta)=\sum_{\mu=1}^{\infty}\scal w_{\mu}^{(0)}\eta^{\mu-1}=\frac{\pi}{\sqrt{2}}\frac{1}{1-\eta}\sqrt{\frac{1+\eta}{1-\eta}}.\label{eq:GenFn}
\end{equation}
Using the coefficients $\scal w_{\mu}^{(0)}$ we can write $\scal{\mat Y}$
symmetrically: defining
\begin{equation}
\mat X(0,\rho)\Def\ee^{-\frac{\rho}{2}\mat{\Phi}_{\Se}\zero}\scal{\mat W}_{\Se}^{(0)}\scal{\mat T}_{\Se,\So}^{(0)}\scal{\mat W}_{\So}^{(0)}\ee^{-\frac{\rho}{2}\mat{\Phi}_{\So}\zero}\label{eq:X_def}
\end{equation}
 we have
\begin{equation}
\scalZ_{\gm}(0,\rho)=\det\left(\mat 1+\bar{\mat X}\mat X\right),\label{eq:Z_of_X}
\end{equation}
where the bar denotes the transpose. As a final remark, we point out
that the general residual matrix (\ref{eq:sc-Y}) can also be written
symmetrically using $\scal w=\sqrt{\scal v}$. However, $\scal v$
from (\ref{eq:sc-v_mu}) is not a formal square. Maybe this symmetric
representation can be utilized to proof the equivalence of (\ref{eq:sc-Z_series})
and (\ref{eq:Sigma(0)}).

From (\ref{eq:Psi2}) and (\ref{eq:sc-Z_series}) the series of the
strip Casimir potential scaling function $\scalF_{\gm}(0,\rho)$ reads,
with $q\Def e^{-2\pi\rho}$, \bS
\begin{align}
\Psi_{\gm}(0,\rho) & =-\log\scalZ_{\gm}(0,\rho)\\
 & =-\frac{1}{4}\left[q+\frac{3}{2}q^{2}+\frac{4}{3}q^{3}+\frac{7}{4}q^{4}+\frac{6}{5}q^{5}+\frac{12}{6}q^{6}+\frac{8}{7}q^{7}+\mathcal{O}(q^{8})\right]\label{eq:sc-F_series3}\\
 & =-\frac{1}{4}\sum_{n=1}^{\infty}\frac{\sigma(n)}{n}q^{n},\label{eq:sc-F_series4}
\end{align}
\eS{}with the divisor sum function $\sigma(n)=\sum_{d|n}d$ \cite{Hardy79}.

The strip Casimir force at $x=0$ is given by \bS
\begin{equation}
\psi_{\gm}(0,\rho)=-\partial_{\rho}\Psi_{\gm}(0,\rho)=-\frac{\pi}{2}\sum_{n=1}^{\infty}\sigma(n)q^{n}=-\frac{\pi}{48}-\frac{\ii}{4}\frac{\eta'(\ii\rho)}{\eta(\ii\rho)}=\frac{\pi}{48}(E_{2}(\ii\rho)-1),\label{eq:psi(0,rho)}
\end{equation}
with Ramanujan\textquoteright s weight two Eisenstein series $E_{2}$
\cite{MathWorld-EisensteinSeries}, and with special value at $\rho=1$,
\begin{equation}
\psi_{\gm}(0,1)=\frac{1}{16}-\frac{\pi}{48}=-0.0029498469\ldots\,.\label{eq:psi(0,1)}
\end{equation}
\eS{}All these results can easily be deduced from different representations
of the double series
\begin{equation}
4\Psi_{\gm}(0,\rho)=-\sum_{j=1}^{\infty}\sum_{k=1}^{\infty}\frac{1}{k}q^{jk}=\sum_{j=1}^{\infty}\log(1-q^{j})=\log(q)_{\infty}=\frac{\pi\rho}{12}+\log\eta(\ii\rho)\label{eq:doubleseries}
\end{equation}
which can be rewritten as a double sum over $n=jk$ and the divisors
$d$ of $n$, 
\begin{equation}
4\Psi_{\gm}(0,\rho)=-\sum_{n=1}^{\infty}\sum_{d|n}\frac{d}{n}q^{n}=-\sum_{n=1}^{\infty}\frac{\sigma(n)}{n}q^{n}.\label{eq:divisorsum}
\end{equation}

The total Casimir force $\vartheta_{\gm}$ at $x_{\gm}=0$ is even
simpler than (\ref{eq:psi(0,rho)}) and reads
\begin{equation}
\vartheta_{\gm}(0,\rho)=\frac{\pi}{48}+\psi_{\gm}(0,\rho)=-\frac{\ii}{4}\frac{\eta'(\ii\rho)}{\eta(\ii\rho)}=\frac{\pi}{48}E_{2}(\ii\rho),\label{eq:vartheta(0,rho)}
\end{equation}
leading to the result of Cardy \& Peschel \cite{CardyPeschel88},
that the amplitude of the logarithmic divergence of the Ising finite
size free energy is $\frac{1}{16}$, as
\begin{equation}
\vartheta_{\gm}(0,1)=\frac{\pi}{48}+\left(\frac{1}{16}-\frac{\pi}{48}\right)=\frac{1}{16}.\label{eq:vartheta(0,1)}
\end{equation}
In chapter \ref{subsec:vartheta} we will return to this point. 

\subsection{Results for general $x$ and $\rho$}

Using (\ref{eq:Sigma-N}), we can calculate the FSS functions for
given $x$ with arbitrary precision, while the aspect ratio $\rho$
remains a free parameter in the expressions. However, we first have
to estimate the surface-corner contribution $\Theta_{\mathrm{s,c}}(x)$
for a complete picture.

\subsection{The Casimir potential scaling function $\Theta(x,\rho)$ \label{subsec:Potential}}

\global\long\def\Li{\mathrm{Li}}
For the computation of the (total) Casimir potential scaling function
$\Theta(x,\rho)$ we need the surface-corner contribution $\Theta_{\mathrm{s,c}}(x)$
from (\ref{eq:Theta-sum}), 
\begin{equation}
\Theta_{\mathrm{s,c}}(x)=-\rho\Thetaoo(x)+\log\scalZ(x,\rho)+\Theta_{\gm}(x_{\gm},\rho)\qquad\forall\rho,\label{eq:Theta_sc(x)}
\end{equation}
with volume scaling variable $x_{\gm}=x\rho^{1/2}$ from (\ref{eq:x-identity}),
that unfortunately could not be calculated directly from the regularized
FSS limit of $\Fscres(\LM)$ yet. However, we can utilize the symmetry
of the square, where $\rho=1$, under the exchange of the two lattice
directions $\L$ and $\M$, which implies $\partial_{\rho}\Theta_{\gm}(x_{\gm},\rho)|_{\rho=1}=0$
\cite{HuchtGruenebergSchmidt11}, together with (\ref{eq:scalrel_gm}),
(\ref{eq:vartheta_def}) and the $2d$ Ising value $d\nu=2$. We get
the scaling relation 
\begin{equation}
\vartheta_{\gm}(x_{\gm},1)=-\frac{x_{\gm}}{2}\frac{\partial}{\partial x_{\gm}}\Theta_{\gm}(x_{\gm},1),\label{eq:vartheta(x,1)-from-Theta(x,1)}
\end{equation}
which can be solved for $\Theta_{\gm}(x_{\gm},1)$ to find, using
(\ref{eq:vartheta-sum}), \bS
\begin{align}
\Theta_{\gm}(x_{\gm},1) & =2\int_{x_{\gm}}^{x_{\gm}\infty}\dd\xi\,\xi^{-1}\vartheta_{\gm}(\xi,1)\label{eq:Theta-DGL-1}\\
 & =\underbrace{-2\int_{x_{\gm}}^{x_{\gm}\infty}\dd\xi\,\xi^{-1}\Thetaoo(\xi)}_{I_{\gm}^{(1)}(x_{\gm})}+\underbrace{2\int_{x_{\gm}}^{x_{\gm}\infty}\dd\xi\,\xi^{-1}\psi_{\gm}(\xi,1)}_{I_{\gm}^{(2)}(x_{\gm})}.\label{eq:Theta-DGL-2}
\end{align}
\eS{}As $\vartheta_{\gm}(0,1)=\frac{1}{16}$, $\Thetaoo(0)=-\frac{\pi}{48}$
and $\psi_{\gm}(0,1)=\frac{1}{16}-\frac{\pi}{48}$ are all finite,
see last chapter, the integrals $I_{\gm}^{(1,2)}(x_{\gm})$, and consequently
$\Theta_{\gm}(x_{\gm},1)$, diverge logarithmically at $x_{\gm}=0$. 

The integral $I_{\gm}^{(1)}(x_{\gm})$ over $\Thetaoo$ from (\ref{eq:Theta-oo})
can be evaluated analytically by exchanging the two integrals and
using the formula $\int_{x}^{\infty}\dd\xi\log(1+a\ee^{-b\xi})=-b^{-1}\Li_{2}(-a\ee^{-bx})$,
with polylogarithm $\Li$, with the result \bS
\begin{equation}
I_{\gm}^{(1)}(x_{\gm})=-\frac{1}{2\pi}\int_{|x_{\gm}|}^{\infty}\dd\Omega\frac{1}{\sqrt{\Omega^{2}-x_{\gm}^{2}}}\Li_{2}\!\left(-\frac{\Omega-x_{\gm}}{\Omega+x_{\gm}}\ee^{-2\Omega}\right).\label{eq:I1(x)}
\end{equation}
To leading order, $I_{\gm}^{(1)}$ diverges logarithmically and has
a jump at zero from the change of the integration limits,
\begin{equation}
I_{\gm}^{(1)}(x_{\gm})\simeq-\frac{\pi}{24}\log|x_{\gm}|-C\,\frac{|x_{\gm}|}{x_{\gm}},\label{eq:I1(x)-simeq}
\end{equation}
\eS{}with Catalan's constant $C$.

The second integral $I_{\gm}^{(2)}(x_{\gm})$ has to be evaluated
numerically, so we split off the log singularity and write \bS
\begin{equation}
I_{\gm}^{(2)}(x_{\gm})=\psi_{\gm}(0,1)\log(1+x_{\gm}^{-2})+2\int_{x_{\gm}}^{x_{\gm}\infty}\dd\xi\,\xi^{-1}\left[\psi_{\gm}(\xi,1)-\frac{1}{1+\xi^{2}}\psi_{\gm}(0,1)\right].\label{eq:I2(x)}
\end{equation}
While the integrand is analytic at $x_{\gm}=0$, the integral again
develops a jump discontinuity at $x_{\gm}=0$ from the different integration
limits above and below zero, 
\begin{equation}
I_{\gm}^{(2)}(x_{\gm})\simeq-\left(\frac{1}{8}-\frac{\pi}{24}\right)\log|x_{\gm}|+\left(C-\frac{3}{4}\log2\right)\frac{|x_{\gm}|}{x_{\gm}}.\label{eq:I2(x)-simeq}
\end{equation}
\eS{}Via the low-temperature integration limit in (\ref{eq:I2(x)})
we have respected the additional contribution $-\log2$ due to the
broken symmetry in the ordered phase, see \cite{HuchtGruenebergSchmidt11}
for details, that was inadvertently attributed to the corner free
energy $\fc^{<}$ in \cite{Baxter16,Hucht16a}. Therefore, $\Theta_{\mathrm{s,c}}(x\to-\infty)=\Theta_{\gm}(x_{\gm}\to-\infty,\rho)=-\log2$
in agreement with the corner-free toroidal and cylindrical case \cite{HuchtGruenebergSchmidt11,HobrechtHucht16a},
while $\fc\to0$ both in the high- and low-temperature limit.

\begin{figure}
\begin{centering}
\includegraphics[scale=1.05]{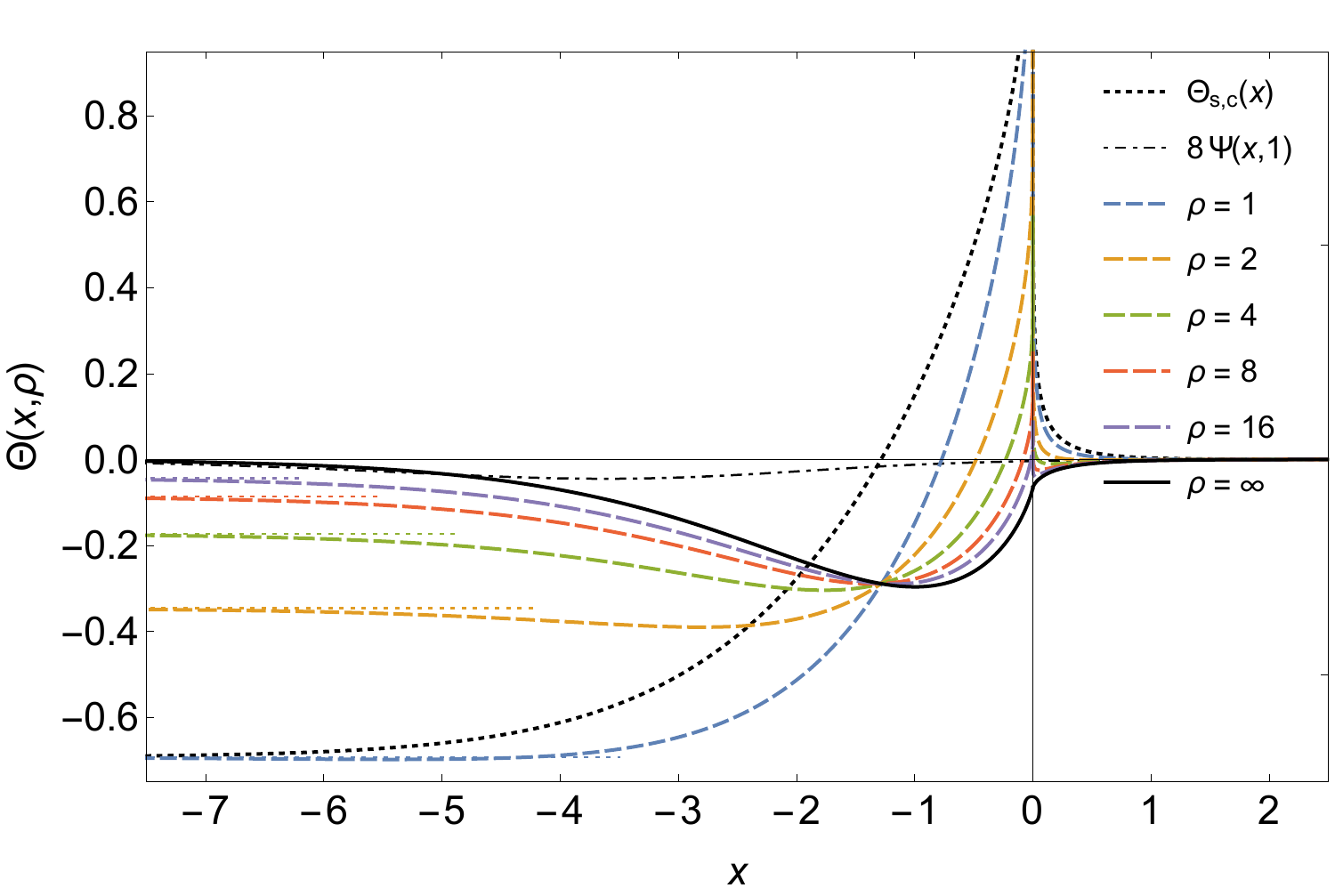}
\par\end{centering}
\caption{Universal Casimir potential scaling function $\Theta(x,\rho)$ for
the Ising rectangle with different aspect ratios $\rho\geq1$, together
with the surface-corner contribution $\Theta_{\mathrm{s,c}}(x)$ and
the strip residual contribution $\Psi(x,1)$. The latter is multiplied
by 8. The corresponding curves for $\rho<0$ fulfill (\ref{eq:Theta_symmetry}).
Note that $\Theta(x\to-\infty,\rho)=-\rho^{-1}\log2$ (dotted lines).
\label{fig:Theta}}
\end{figure}
As both $\Thetaoo(x)$ and $\log\scalZ(x,\rho)$ are analytic around
$x=0$, $\Theta_{\mathrm{s,c}}(x)$ from (\ref{eq:Theta_sc(x)}) fulfills
\begin{equation}
\Theta_{\mathrm{s,c}}(x)=-\frac{1}{8}\log|x|-\frac{3}{4}\log2\,\frac{|x|}{x}+\textit{regular terms}\label{eq:Theta_s,c-simeq}
\end{equation}
for small $x$. The same asymptotic behavior holds for the Casimir
potential, but with a different (in)dependency on $\rho$, 
\begin{equation}
\Theta_{\gm}(x_{\gm},\rho)=-\frac{1}{8}\log|x_{\gm}|-\frac{3}{4}\log2\,\frac{|x_{\gm}|}{x_{\gm}}+\textit{terms regular in }x_{\gm}.\label{eq:Theta_o-simeq}
\end{equation}
Remember that $\Theta_{\gm}$ is a function symmetric under $\L/\M$
exchange, while $\Theta_{\mathrm{s,c}}$ is a property of the $\L$
boundary. The result for $\Theta_{\mathrm{s,c}}(x)$ is shown in figure
\ref{fig:Theta}, together with the Casimir potential scaling function
$\Theta(x,\rho)$ for several values of $\rho\geq1$. The behavior
of both quantities is dominated by the logarithmic divergence at $x=0$.
Furthermore, both are much larger below criticality, which is expected
for systems with open boundary conditions.

While the (total) Casimir potential $\Theta(x,\rho)$ diverges logarithmically
for $x\to0$ and therefore is infinite at criticality for all finite
aspect ratios $0<\rho<\infty$, we can nevertheless subtract the divergent
contribution $\Theta_{\mathrm{s,c}}(x)$ and define a finite Casimir
amplitude 
\begin{equation}
\Delta_{\gm}(\rho)\Def\lim_{x\to0}[\rho\Theta(x,\rho)-\Theta_{\mathrm{s,c}}(x)]=\rho\Thetaoo(0)+\Psi_{\gm}(0,\rho)=\frac{1}{4}\log\eta(\ii\rho).\label{eq:Delta_gm}
\end{equation}
As consequence, we have the unusual situation that the (finite) Casimir
amplitude is different from the (divergent) Casimir potential at criticality.

In summary, we observe the following logarithmic contributions to
the total free energy $F$ and its residual $\Fres$: the total finite-size
free energy (\ref{eq:F}) has a log term at criticality, originally
predicted by Cardy \& Peschel \cite{CardyPeschel88} using conformal
field theory, \bS
\begin{equation}
F(\tau{=}0;L,M)=LM\fb(0)+[L+M]\fs(0)-\frac{1}{16}\log(LM)+\textit{regular terms},\label{eq:F-simeq}
\end{equation}
with reduced temperature $\tau=1-z/\zc$, while the infinite volume
contribution (\ref{eq:F_inf}) has log terms and a jump originating
from the corner free energy, see (\ref{eq:fc-expansion}) in the appendix,
\begin{equation}
F_{\infty}(\tau;L,M)=LM\fb(\tau)+[L+M]\fs(\tau)+\frac{1}{8}\log|\tau|+\frac{3}{4}\log2\,\frac{|\tau|}{\tau}+\textit{regular terms}\label{eq:F_inf-simeq}
\end{equation}
such that
\begin{equation}
\Fres\left(\tau;L,M\right) = -\frac{1}{16}\log(\tau^{2}LM)-\frac{3}{4}\log2\,\frac{|\tau|}{\tau}+\textit{regular terms}.\label{eq:F_inf^res-simeq}
\end{equation}
\eS{}In the FSS limit, where $x_{\gm}=2\tau\sqrt{LM}$ and $\rho=L/M$,
both log contributions combine to (\ref{eq:Theta_o-simeq}). We conclude
that both the logarithmic divergence and the jump in the Casimir potential
scaling function $\Theta_{\gm}(x_{\gm},\rho)$ (\ref{eq:Theta_o-simeq})
stem from the near critical behavior of the corner free energy $\fc$.

\subsection{The Casimir force scaling function $\vartheta(x,\rho)$ \label{subsec:vartheta}}

\begin{figure}
\begin{centering}
\includegraphics[scale=1.05]{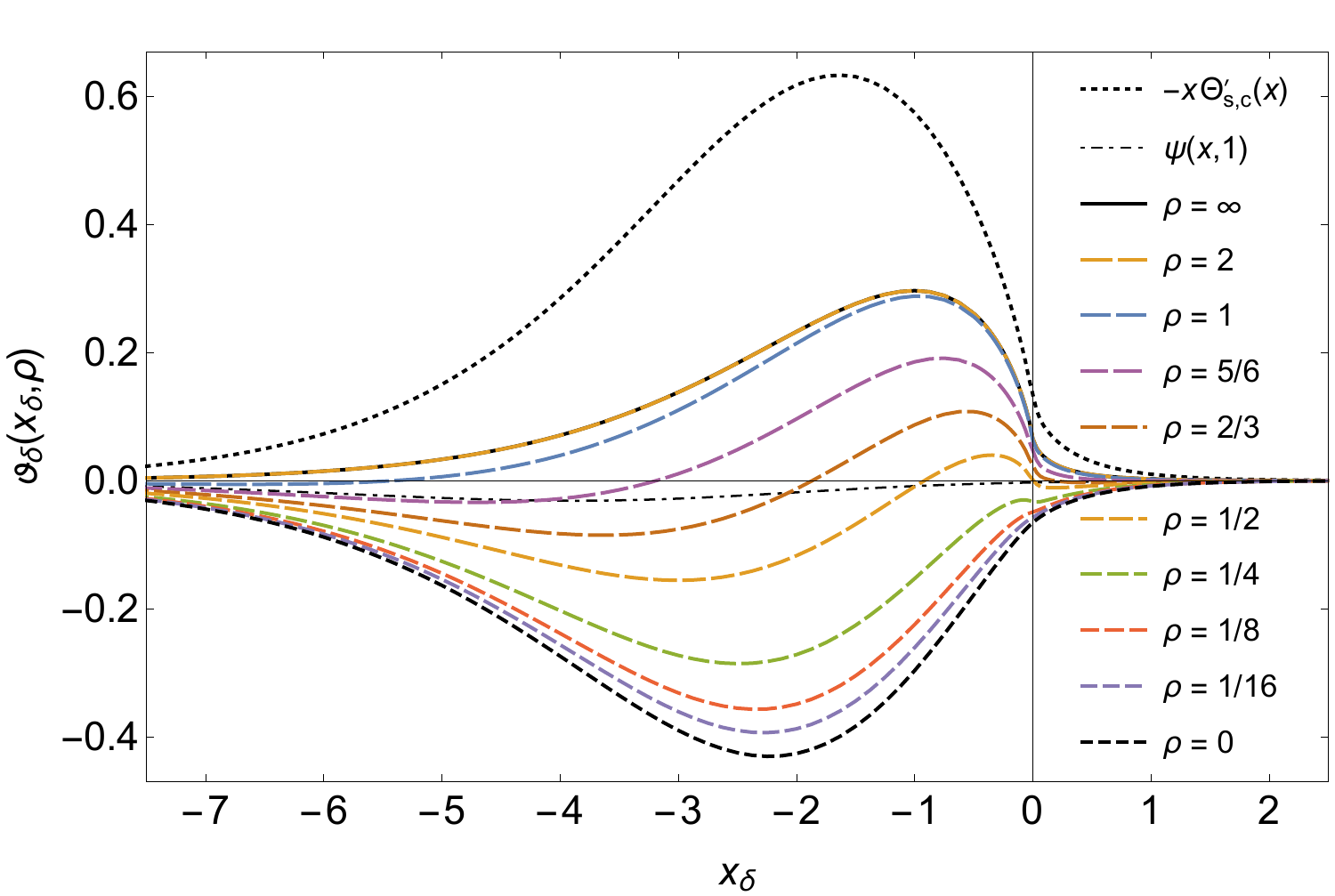}
\par\end{centering}
\caption{Universal Casimir force scaling function $\vartheta_{\delta}(x_{\delta},\rho)$
for the Ising rectangle with different aspect ratios $\rho$. For
$\rho\geq1$ we show $\vartheta\protect\Def\vartheta_{\protect\M}$
over $x\protect\Def\protect\xM$ as defined by (\ref{eq:scalrel_para}),
while for $\rho\leq1$ we show $\vartheta_{\protect\L}$ over $\protect\xL$
from (\ref{eq:scalrel_perp}). Note that the curve for $\rho=\infty$
is masked behind the curve for $\rho=2$. Also shown are the surface-corner
contribution $-x\Theta'_{\mathrm{s,c}}(x)$ from (\ref{eq:xTheta_sc})
and the strip contribution $\psi(x,1)$ from (\ref{eq:FC_strip-scal})
to the Casimir force. \label{fig:vartheta}}
\end{figure}

Finally we come to the Casimir force scaling function $\vartheta(x,\rho)$
at arbitrary $x$ and $\rho$. From (\ref{eq:vartheta-sum}) we can
easily determine the Casimir force with high precision, provided $\rho\gtrsim1$,
\begin{equation}
\vartheta(x,\rho)=-\Theta^{(\mathrm{oo})}(x)+\psi(x,\rho).\label{eq:vartheta-sum1}
\end{equation}
For $\rho\lesssim1$, however, the convergence is suboptimal, and
we instead use (\ref{eq:scalrel_perp}) and the $\L,\M$ exchange
symmetry and calculate the Casimir force in $\M$ direction instead,
to get the equivalent expression 
\begin{equation}
\vartheta(x,\rho)=\vartheta^{(\mathrm{oo})}(x)-\rho x\Theta'_{\mathrm{s,c}}(x)-\frac{x\partial}{\partial x}\Psi(x,\rho^{-1})-\psi(x,\rho^{-1}),\label{eq:vartheta-rho-small}
\end{equation}
with \cite{Au-YangFisher80,EvansStecki94,BrankovDantchevTonchev00}
\begin{equation}
\vartheta^{(\mathrm{oo})}(x)\Def-\frac{1}{\pi}\int_{0}^{\infty}\dd\omega\,\sqrt{x^{2}+\omega^{2}}\left(1+\frac{\sqrt{x^{2}+\omega^{2}}+x}{\sqrt{x^{2}+\omega^{2}}-x}\ee^{2\sqrt{x^{2}+\omega^{2}}}\right)^{-1}\label{eq:theta-oo}
\end{equation}
from (\ref{eq:scalrel_perp}), that does not suffer from convergence
problems if $\rho\lesssim1$. At $\rho=1$ we can derive the expression
\begin{equation}
x\Theta'_{\mathrm{s,c}}(x)=\Thetaoo(x)+\vartheta^{(\mathrm{oo})}(x)-2\psi(x,1)-\frac{x\partial}{\partial x}\Psi(x,1),\label{eq:xTheta_sc}
\end{equation}
that implies that $-x\Theta'_{\mathrm{s,c}}(x)$ is approximately
equal to the difference $\vartheta_{\M}(x_{\M},\infty)-\vartheta_{\L}(x_{\L},0)$,
that is, the distance between the dashed and the solid black curves
in figure \ref{fig:vartheta}.

The resulting universal Casimir force scaling functions for different
values of the aspect ratio $\rho$ are displayed in figure \ref{fig:vartheta}.
The force is attractive for small aspect ratios $\rho\lesssim1/4$
and becomes repulsive for larger aspect ratios, a behavior which is
very similar to the fully periodic case \cite{HuchtGruenebergSchmidt11,HobrechtHucht16a}.
However, the force at criticality $x=0$ in a square system $\rho=1$
does not vanish such as in the periodic case. This is directly related
to the log divergence of the corresponding Casimir potential and is
interpreted as a consequence of a long-range repulsive corner-corner
interaction. At criticality, the Casimir force changes sign at $\rho_{\text{0}}=0.523521700017999266800\ldots$
and is attractive for $\rho<\rho_{0}$. Note that $\ii\rho_{0}$ is
the only purely imaginary zero of the Eisenstein series $E_{2}$.

\newpage{}

\section{Scaling limit of effective spin model}

Finally, we discuss the FSS limit of the effective spin model introduced
in \cI{99}. In the FSS limit $M\to\infty$, $T\to\Tc$ with constant
$x$ and $\rho$, we find the thermodynamic limit $N\to\infty$ of
the scaled reduced Hamiltonian 
\begin{equation}
\scal{\mathcal{H}}_{\mathrm{eff}}(x,\rho)=-\sum_{\mu<\nu=1}^{N}\scal K_{\mu\nu}(x)s_{\mu}s_{\nu}+\rho\sum_{\mu=1}^{N}\Gamma_{\mu}(x)s_{\mu}+b\Big[\sum_{\mu=1}^{N}\sigma_{\mu}s_{\mu}\Big]^{2},\label{eq:sc-H_eff}
\end{equation}
with scaled interaction constants 
\begin{equation}
\scal K_{\mu\nu}=-\sigma_{\mu}\sigma_{\nu}\log\frac{\scal v_{\mu}\scal v_{\nu}}{(\Phi_{\mu}^{2}-\Phi_{\nu}^{2})^{2}},\label{eq:sc-K_munu}
\end{equation}
and with $\scal v_{\mu}$ from (\ref{eq:sc-v_mu}).

For large $\mu+\nu$, with $\nu-\mu\ll\nu+\mu$, the interactions
are asymptotically
\begin{equation}
\scal K_{\mu\nu}\simeq2\sigma_{\mu}\sigma_{\nu}\log\left[\frac{\pi}{2}|\nu-\mu|\right],\label{eq:sc-K_munu-1}
\end{equation}
while for large $\nu$ with $\mu\ll\nu$ 
\begin{equation}
\scal K_{\mu\nu}=2\sigma_{\mu}\sigma_{\nu}\log\left[\frac{\pi^{3/2}\nu^{3/2}}{2^{3/2}(\mu-\frac{1}{2})\Beta(\frac{\mu}{2},\frac{1}{2})}[1+\mathcal{O}(\nu^{-1})]\right].\label{eq:sc-K_munu-1-1}
\end{equation}
In all cases the interaction grows logarithmically with $\mu$ and
$\nu$. 

The aspect ratio $\rho$ takes the role of an applied homogeneous
magnetic field, which acts on the spins $s_{\mu}\in\{0,1\}$ via a
magnetic moment $\Gamma_{\mu}$ that grows linearly with $\mu$. As
a consequence, the spins are asymptotically fixed to $s_{\mu}=0$
for large $\mu$ if $\rho>0$, such that the spin dynamics is mainly
restricted to the first few spins.

The strip Casimir force scaling function is related to the magnetization
scaling function of the effective model, c.f. \cI{103},
\begin{equation}
\psi(x,\rho)=\frac{\partial}{\partial\rho}\log\scalZ(x,\rho)=-\Big\langle\sum_{\mu=1}^{\infty}\Gamma_{\mu}s_{\mu}\Big\rangle_{\mathrm{eff}}=-\sum_{\mu=1}^{\infty}\Gamma_{\mu}\langle s_{\mu}\rangle_{\mathrm{eff}}=-\scal M_{\mathrm{eff}}(x,\rho).\label{eq:tildeM}
\end{equation}
To summarize, we can define a universal effective spin model (\ref{eq:sc-H_eff})
describing the FSS limit of the $2d$ Ising model on the rectangle.
The couplings $\scal K_{\mu\nu}$ between the spins as well as the
magnetic moments $\Gamma_{\mu}$ depend on $x$, while the aspect
ratio $\rho$ takes the role of a homogeneous magnetic field. Thermodynamic
quantities of this model are directly related to universal scaling
functions of the underlying Ising universality class.

\section{Conclusions}

Based on the results published recently \cite{Hucht16a}, we calculated
the universal finite-size scaling functions of the Casimir potential
and the Casimir force for the Ising universality class on the $L\times M$
rectangle, with open boundary conditions in both directions, and with
arbitrary aspect ratio $\rho$. The calculations were done in the
finite-size scaling limit $L,M\to\infty$, $T\to\Tc$, with fixed
temperature scaling variable $x\propto(T/\Tc-1)M$ and fixed aspect
ratio $\rho\propto L/M$. We have analytically derived exponentially
fast converging series for the related Casimir potential and Casimir
force scaling functions. At the critical point $T=\Tc$ we could confirm
predictions from conformal field theory for both the size dependent
critical free energy (\ref{eq:F-simeq}) \cite{CardyPeschel88} as
well as for the shape dependence of the Casimir amplitude $\Delta_{\gm}(\rho)$
(\ref{eq:Delta_gm}) \cite{KlebanVassileva91}. 

The presence of corners and the related corner free energy has dramatic
impact on the Casimir scaling functions and leads to a logarithmic
divergence of the Casimir potential scaling function at criticality.
As consequence, we have the unusual situation that the (finite) Casimir
amplitude is different from the (divergent) Casimir potential at criticality.
This behavior was not known from other geometries and boundary conditions. 

These strong influence of system corners on the critical Casimir force
might give rise to new applications in the framework of colloidal
suspensions confined in finite near-critical binary liquid mixtures
\cite{HobrechtHucht14,HobrechtHucht15a}.
\begin{acknowledgments}
The author thanks Hendrik Hobrecht, Felix M. Schmidt and Jesper L.
Jacobsen for helpful discussions and inspirations. I also thank my
wife Kristine for her great patience during the preparation of this
manuscript. This work was partially supported by the Deutsche Forschungsgemeinschaft
through Grant HU~2303/1-1.
\end{acknowledgments}

\appendix

\section{Series expansion for zeroes $\Phi_{\mu}$}

\global\long\def\Pz{\Phi_{0,\mu}}
\global\long\def\dPhi{\Delta_{\mu}}
We derive a series representation of $\Phi_{\mu}$ in terms of the
values $\Phi_{\mu}^{(0)}$ at $x=0$ around $\Phi_{\mu}^{(0)}=\infty$:
writing $\Pz\Def\Phi_{\mu}^{(0)}$ and with
\begin{equation}
\Phi_{\mu}=\Pz+\dPhi\label{eq:Phi_with_Delta}
\end{equation}
we have
\begin{equation}
P(\Phi_{\mu})=P(\Pz+\dPhi)=\frac{x}{\Pz+\dPhi}\cos\dPhi-\sin\dPhi=0,\label{eq:P(Phi)_with_Delta}
\end{equation}
leading to the recursion relation
\begin{equation}
\dPhi^{(k)}\mapsto\dPhi^{(k+1)}=\arctan\left(\frac{x}{\Pz+\dPhi^{(k)}}\right).\label{eq:dPhi_recursion}
\end{equation}
This recursion can easily be performed analytically with a computer
algebra system like \emph{Mathematica} \cite{MMA11}: Starting with
an empty series expansion $\dPhi^{(0)}=\mathcal{O}(\Pz^{-1})$, we
can simply apply (\ref{eq:dPhi_recursion}) $n$ times using the command
\begin{equation}
\mathtt{\Delta_{\mu}[n\_]:=Nest\big[ArcTan\Big[\frac{x}{\Phi_{0,\mu}+\#}\Big]\&,Series\Big[\frac{1}{\Phi_{0,\mu}},\{\Phi_{0,\mu},\infty,0\}\Big],n\big]}\label{code:recursion}
\end{equation}
to get the correct series expansion up to $\mathcal{O}(\Pz^{-(2n+1)})$,
with the result
\begin{equation}
\Phi_{\mu}^{2}=\Pz^{2}+2x-\frac{x^{2}(2x+3)}{3\Pz^{2}}+\frac{2x^{3}(x^{2}+5x+5)}{5\Pz^{4}}+\mathcal{O}(\Pz^{-6})\label{eq:Phi_series}
\end{equation}
for $\Phi_{\mu}^{2}$.

\section{Expansion of $q$-products around $q=1$ \label{sec:q-expansion}}

The isotropic corner free energy near $\Tc$ can be derived from the
$q$-product representation of Vernier \& Jacobsen \cite{VernierJacobsen12}
as well as from \cI{A7d}. Written in terms of the reduced temperature
$\tau=1-z/\zc$, the expansion is given by 
\begin{equation}
\fc(\tau)=\frac{1}{8}\log|\tau|-\frac{2}{\pi}C+\frac{9}{16}\log2+\frac{3}{4}\log2\,\frac{|\tau|}{\tau}+\mathcal{O}(\tau),\label{eq:fc-expansion}
\end{equation}
 with Catalan's constant $C$.

The isotropic surface free energy near $\Tc$ is calculated using
the result of McCoy \& Wu \cite[(4.35)]{McCoyWu73} together with
an expansion of the $q$-products derived in \cite{VernierJacobsen12}
and \cI{A7} to be 
\begin{equation}
f_{\mathrm{s}}(\tau)=f_{\mathrm{s}}(0)+\frac{|\tau|}{2}+\left(\frac{1}{4}-\frac{3\log2}{2\pi}+\frac{\log|\tau|-1}{\pi}\right)\tau+\mathcal{O}(\tau^{2}).\label{eq:fs-expansion}
\end{equation}
 The critical value reads
\begin{align}
f_{\mathrm{s}}(0) & =-\frac{3}{4}\log\zc-2\left[\zeta^{(1,0)}(-1,{\textstyle \frac{1}{8}})+\zeta^{(1,0)}(-1,{\textstyle \frac{3}{8}})-\zeta^{(1,0)}(-1,{\textstyle \frac{5}{8}})-\zeta^{(1,0)}(-1,{\textstyle \frac{7}{8}})\right]\label{eq:fs-critical}\\
 & =0.1817314169844\ldots,\nonumber 
\end{align}
with generalized Riemann zeta function $\zeta(s,a)=\sum_{k=0}^{\infty}(k+a)^{-s}$.
Note that the exact value of the critical surface free energy $\fs(0)$
given in (\ref{eq:fs-critical}) was not published yet. Both (\ref{eq:fc-expansion})
and (\ref{eq:fs-critical}) can be derived from the product representations
\cI{A5b} and \cI{A7b}, expanded around the limit $q\to1$.

\bibliographystyle{unsrt}
\phantomsection\addcontentsline{toc}{section}{\refname}\bibliography{Physik}

\end{document}